\pdfoutput=1
\RequirePackage{ifpdf}
\ifpdf 
\documentclass[pdftex]{sigma}
\else
\documentclass{sigma}
\fi

\DeclareMathOperator{\arcsinh}{arcsinh}
\DeclareMathOperator{\arccosh}{arccosh}

\numberwithin{equation}{section}

\begin{document}

\allowdisplaybreaks

\renewcommand{\thefootnote}{$\star$}

\newcommand{\arXivNumber}{1504.06228}

\renewcommand{\PaperNumber}{096}

\FirstPageHeading

\ShortArticleName{Harmonic Oscillator on the ${\rm SO}(2,2)$ Hyperboloid}

\ArticleName{Harmonic Oscillator on the $\boldsymbol{{\rm SO}(2,2)}$ Hyperboloid\footnote{This paper is a~contribution to the Special Issue
on Analytical Mechanics and Dif\/ferential Geometry in honour of Sergio Benenti.
The full collection is available at \href{http://www.emis.de/journals/SIGMA/Benenti.html}{http://www.emis.de/journals/SIGMA/Benenti.html}}}

\Author{Davit R.~{PETROSYAN}~$^\dag$ and George S.~{POGOSYAN}~$^{\ddag\S}$}

\AuthorNameForHeading{D.R.~Petrosyan and G.S.~Pogosyan}

\Address{$^\dag$~Laboratory of Theoretical Physics, Joint Institute for Nuclear Research,\\
\hphantom{$^\dag$}~Dubna, Moscow Region, 141980, Russia}
\EmailD{\href{mailto:petrosyan@theor.jinr.ru}{petrosyan@theor.jinr.ru}}

\Address{$^\ddag$~Departamento de Matematicas, CUCEI, Universidad de Guadalajara,\\
\hphantom{$^\ddag$}~Guadalajara, Jalisco, Mexico}
\EmailD{\href{mailto:george.pogosyan@cucei.udg.mx}{george.pogosyan@cucei.udg.mx}}

\Address{$^\S$~International Center for Advanced Studies, Yerevan State University,\\
\hphantom{$^\S$}~A.~Manoogian 1, Yerevan, 0025, Armenia}
\EmailD{\href{mailto:pogosyan@ysu.am}{pogosyan@ysu.am}}

\ArticleDates{Received April 24, 2015, in f\/inal form November 20, 2015; Published online November 25, 2015}

\Abstract{In the present work the classical problem of harmonic oscillator in the hyperbolic space $H_2^2$: $z_0^2+z_1^2-z_2^2-z_3^2=R^2$
has been completely solved in framework of Hamilton--Jacobi equation. We have shown that the harmonic oscillator on $H_2^2$, as in the other
spaces with constant curvature, is exactly solvable and belongs to the class of maximally superintegrable system.
We have proved that all the bounded classical trajectories are closed and periodic. The orbits of motion are ellipses or circles for bounded
motion and ultraellipses or equidistant curve for inf\/inite ones.}


\Classification{22E60; 37J15; 37J50; 70H20}

\renewcommand{\thefootnote}{\arabic{footnote}}
\setcounter{footnote}{0}

\section{Introduction}

The harmonic oscillator as a distinguished dynamical system plays the fundamental role in theoretical and mathematical physics
due to many special properties outgoing from its hidden symmetry. Together with the Kepler--Coulomb problem they are only one
among the central potentials for which all classical trajectories are closed (Bertrand theorem) and in quantum mechanics all energy 
state are multiply degenerate (accidental degeneracy). The other consequence of hidden symmetry is the existence of
additional functionally (in quantum mechanics linearly) independent integrals of motion and the phenomena of multiseparability, that is 
separability of variables in Hamilton--Jacobi or Schr\"odinger equation in more than one orthogonal systems of coordinate. 
It has long been known~\cite{DEMKOV,FRADK, JAUHILL} that the harmonic oscillator problem possesses f\/ive functionally
independent
integrals of motion, which generate the separation of variable into eight systems of coordinates~\cite{EVANS,GROP1}. 
In most of them harmonic oscillator admits the exact solution, the fact which makes it attractive to use as a model of molecular, 
atomic and nuclear physics and other branches of theoretical physics. 

The generalization of Kepler--Coulomb system and oscillator problem on the spaces of constant curvature start from the work of 
Lobachevsky, who f\/irst identif\/ied the Kepler potential in hyperbolic space $H_3$ (two-sheeted hyperboloid)
and found the trajectories of classical
motion~\cite{LOBACHEVSKI} (see also the articles \cite{chern,DOMBROWSKI,KOZLOV1,SLAW}).  The extension of the
harmonic oscillator 
problem on the spherical and hyperbolic geometries has  already been done in the book of Liebmann~\cite{LIEBMANN}, who
also discussed
the geometric character of the conics in noneuclidean geometry. The investigation of Kepler--Coulomb problem in quantum
mechanics
was motivated to compare the properties of the Coulomb potential in the ``open hyperbolic'' or ``closed'' universe to that of an 
``open but f\/lat'' universe. Schr\"odinger \cite{SCHRO} was the f\/irst who discussed this problem and discovered that for ``hydrogen atom''
on three-dimensional sphere only discrete spectrum exists. Virtually at the same time, Infeld and Shild~\cite{INFELD} found that 
in an open hyperbolic universe there is only a f\/inite (but very large) number of bound states. The motion in Coulomb f\/ield on 
imaginary Lobachevsky space (one-sheet hyperboloid), as shown by Grosche~\cite{GROaa}, has some peculiarities.
It is not singular for any value of variable and their discrete spectra inf\/inite degenerate. The essential advance in the
theory of systems with
hidden symmetry in the spaces with constant curvature was made by Higgs~\cite{HIG}, Leemon~\cite{LEEMON} and Belorussian authors
in~\cite{BOKUOT}. They have shown that the complete degeneracy of the spectrum of the Coulomb and oscillator problems on the three-
dimensional sphere and hyperboloid is caused by an additional integrals of motion: ``curved'' Runge--Lenz's vector (for the Coulomb potential)
and Demkov--Fradkin tensor (for the oscillator). However, in contrast to the f\/lat space, commutation relations between the components of  
Runge--Lenz's vector and Demkov--Fradkin tensor on the sphere and hyperboloid form the quadratic or cubic algebra. Later it was proven
that these properties are inherent in all class of  maximally second-order superintegrable systems, which also belong to the Kepler--Coulomb
and oscillator potentials (see for instance recent review~\cite{MPW} and references therein). 

We recall that in general, in an $N$-dimensional space, maximal superintegrability means that the classical Hamiltonian allows
$(2N-1)$ functionally independent integrals of motion (including the Hamiltonian) that are well def\/ined functions on phase space. The f\/irst
search of superintegrable systems in two- and three-dimensional f\/lat space was done in the pioneering works of Winternitz and
Smorodinsky with co-authors in~\cite{WS2, WS1}, later the notion of superintegrability in the spaces of constant curvature has been
introduced in the series of papers \cite{GROP1,GROP5,GROP0,GROPOSI}. The complete classif\/ication of superintegrable systems on the
two-dimensional complex sphere, which include to real spaces, sphere and hyperboloid, as particular cases have done in the work~\cite{KKMP}.
Some of the superintegrable systems have been constructed on~$S_N$ and~$H_N$ spaces in~\cite{HERBAL1}. We can also mention some articles
devoted to the investigation of various aspects of both classical and quantum superintegrable systems in the spaces of constant curvature,
for instance~\cite{BOKUOT,HAKPOG,KAL-MIL,KMP5,KMP4,POG1}. 

The classical and quantum mechanical systems on the spaces of constant curvature (positive and negative) have always drawn
a great attention due to their  connection with the relativistic physics and gravity. The 2D and 3D one-sheeted and ${\rm SO}(2,2)$
hyperboloids are the models of the relativistic space time with a constant curvature, namely de Sitter and anti de Sitter spaces,
which is a crucial point for its wide application in the f\/ield theories~\cite{PARKER,Varlamov}, quantum gravity and
cosmology~\cite{ Ambrozio,Gibbons,HOOFT}, integrable Yang--Mills--Higgs equation (or Bogomolny equation)~\cite{WARD,ZHOU}.
Among other applications we can mention also quantum Hall ef\/fect~\cite{HALL} and coherent state quantization~\cite{GAZEAU}. 

However, as far as we know, the superintegrable systems on imaginary Lobachevski space $H_2^1$: ${\rm SO}(3,1) / {\rm SO}(2,1)$, 
(de Sitter space time ${\bf dS}^{2+1}$) on hyperboloid $H_2^2 = {\rm SO}(2,2) / {\rm SO}(2,1)$, (Anti de Sitter space time {\bf AdS}$^{2+1}$), 
have not been studied with the same degree of detail and need to be further investigated. It appears that the f\/irst work
in this direction
(if we do not take into account the paper~\cite{GROaa}) was the article~\cite{OLMO} (see also more general case in \cite{calz})
where the authors, using the reduction procedure to the free Hamiltonian on the homogeneous space ${\rm SU}(2,2)/{\rm U}(2,1)$, obtain the eleven dif\/ferent
types of maximally superintegrable systems on the hyperboloid~$H_2^2$.  Later, in paper \cite{HERRBALL}, 
the superintegrable generalization of harmonic oscillator and Kepler--Coulomb potentials covering 
the six three-dimensional spaces of constant curvature (including de Sitter and anti de Sitter spaces) in unif\/ied way,
parametrized by two contraction parameters def\/ining the metric in each space, have been constructed.
In these papers the classical superintegrable systems are only identif\/ied but have not been solved.
Recently, also the main properties of two-dimensional harmonic oscillator problem have been
investigated in~\cite{RANADA1}, using again two parameters approach, in nine standard two-dimensional Cayley--Klein spaces,
including the de Sitter ${\bf dS}^{1+1}$ and anti de Sitter ${\bf AdS}^{1+1}$ spaces.       

\looseness=-1
The present work in a sense can be considered as a continuation of our previous ar\-tic\-les~\cite{PETPOG1,PETPOG2,PETPOG3}, 
devoted to the investigation of classical and quantum Kepler--Coulomb problem and quantum harmonic oscillator problem 
on the conf\/iguration hyperbolic space with constant curvature~$H_2^2$.  The given paper aims to investigate the harmonic oscillator problem
on the whole hyperbolic space~$H_2^2$ from the point of view of classical mechanics, which, to our know\-led\-ge, has not been elucidated in literature
so far. This task seems more complicated but also more interesting than the analogous problem in the other three-dimensional hyperbolic spaces.
It mainly derive from the complexity of the space~$H_2^2$  which includes such subspaces as the one- and two-sheeted hyperboloids.
This study will hopefully also help us to better understand the quantum case.

\section[The hyperbolic space $H_2^2$ and constants of motion]{The hyperbolic space $\boldsymbol{H_2^2}$ and constants of motion}

A three-dimensional hyperboloid  $H_2^2$$\subset${\bf R}$_{2,2}$ is described by the equation
\begin{gather}
\label{hyper-01}
z_0^2 + z_1^2 - z_2^2 - z_3^2 = R^2
\end{gather}
To be more specif\/ic we parametrize the hyperboloid (\ref{hyper-01}) using the geodesic pseudo-spherical coordinate
($r, \tau, \varphi$)~\cite{KALMIL1, PETPOG1}, namely
\begin{alignat}{3}
& z_0= \pm R\cosh r,\qquad && z_1=R\sinh r\sinh\tau,& \nonumber\\
& z_2=R\sinh r\cosh\tau\cos\varphi, \qquad && z_3=R\sinh r\cosh\tau\sin\varphi,& \label{coor:1}
\end{alignat}
where $r \geq 0$ is the ``geodesic radial angle'', $\tau \in (-\infty, \infty)$, and $\varphi \in [0, 2\pi)$.
The connection between two sets of coordinates $z_0 \to - z_0$ corresponds to the complex transformation of radial
angle $r \to i\pi - r$.
The system of coordinate~(\ref{coor:1}) is valid only for~$|z_0| \geq R$ and the missing part of the surface
for $|z_0| < R$ may also be taken into account if we use another form of the pseudo-spherical  coordinate
\begin{alignat}{3}
& z_0= \pm R\cos\chi,\qquad && z_1= R\sin\chi\cosh\mu,& \nonumber\\
& z_2=R\sin\chi\sinh\mu\cos\varphi, \qquad && z_3=R\sin\chi\sinh\mu\sin\varphi,& \label{PSEUDO:2}
\end{alignat}
where now $\chi \in (-\frac{\pi}{2}, \frac{\pi}{2})$, $\mu \in (-\infty, \infty)$ and $\varphi \in [0, 2\pi)$.
It is also easy to  see that the two pseudo-spherical system of coordinate~(\ref{coor:1}) and~(\ref{PSEUDO:2})
are connected by
\begin{gather}
\label{TRANSO-01}
r \to i\chi,
\qquad
\tau \to \mu - i\pi/2.
\end{gather}
Here we shall make use of the pseudo-spherical system of coordinate in form~(\ref{coor:1}).
To investigate the motion in the region $|z_0| \leq R$, everywhere below, we will use the
transformation~(\ref{TRANSO-01}).

The restriction of the pseudo-euclidean metric $ds^2 = G_{\mu \nu} d z^{\mu} d z^{\nu}$,
$G_{\mu \nu} = \operatorname{diag} (-1, -1, 1, 1)$, ($\mu, \nu = 0,1,2,3$) on ${\bf R}_{2,2}$ to $H_2^2$
leads to the following formula
\begin{gather*}
\frac{ds^2}{R^2} = d r^2 - \sinh^2 r d \tau^2 + \sinh^2 r \cosh^2\tau d\varphi^2.
\end{gather*}
Then the kinetic energy is given by
\begin{gather*}
{\cal T}  = \frac{R^2}{2}\big(\dot{r}^2 - \sinh^2 r \big(\dot{\tau}^2 -  \cosh^2\tau \dot{\varphi}^2\big)\big)
\end{gather*}
and the canonical momenta can be obtained in a usual way
\begin{gather*}
p_r = \frac{\partial{\cal T}}{\partial\dot{r}} = R^2 \dot{r},
\qquad
p_\tau = \frac{\partial{\cal T}}{\partial\dot{\tau}} = - R^2\sinh^2 r \dot{\tau},
\qquad
p_\varphi = \frac{\partial{\cal T}}{\partial\dot{\varphi}} = R^2 \sinh^2 r \cosh^2 \tau \dot{\varphi}.
\end{gather*}
Thus the free Hamiltonian in the pseudo-spherical phase space $(r,\tau,\varphi; p_r, p_\tau, p_\varphi)$
with respect to the canonical Lie--Poisson brackets
\begin{gather}
\label{CANONIC-01}
\{f, g \}  = \sum_{i=1}^{3}\left(\frac{\partial f}{\partial q_i}\frac{\partial g}{\partial p_i}
- \frac{\partial g}{\partial q_i}\frac{\partial f}{\partial p_i} \right),
\end{gather}
has the form
\begin{gather}
\label{LAGRANGE-04}
{\cal H}_{\rm free}  = \frac{1}{2 R^2}\left\{p_r^2 - \frac{1}{\sinh^2 r}\left(p_\tau^2 -  \frac{p_\varphi^2}{\cosh^2\tau}\right)\right\}.
\end{gather}

It is clear that isometry group of $H_2^2$ hyperboloid is given by ${\rm SO}(2,2)$ group. The correspon\-ding Lie algebra is six dimensional.
The generators of ${\mathfrak{so}}(2,2)$ algebra can be written in terms of the ambient space ${\bf R}_{2,2}$ coordinates~$z_\mu$ and momenta~$p_\mu$ as
\begin{alignat}{4}
& {\cal L}_1  = - (z_2 p_3 - z_3 p_2),  \qquad &&{\cal L}_2  = - (z_1 p_3 + z_3 p_1),  \qquad &&   {\cal L}_3  =  (z_1 p_2 + z_2 p_1),&
\nonumber\\
& {\cal N}_1  = (z_0 p_1 - z_1 p_0),   \qquad &&{\cal N}_2  = - (z_0 p_2 + z_2 p_0),  \qquad &&  {\cal N}_3   =  - (z_0 p_3 + z_3 p_0), &\label{classical1}
\end{alignat}
and the Lie--Poisson brackets (\ref{CANONIC-01}) with the help of three-dimensional metric
$\bar{g}_{ik}\!= \!\operatorname{diag}\{1,\!{-}1,\!{-}1\}$ reads
\begin{gather*}
\{{\cal L}_i,{\cal L}_j\} = \bar{g}_{im}\bar{g}_{jn}\varepsilon_{mnk} {\cal L}_k,
\qquad
\{{\cal N}_i, {\cal N}_j\}= \bar{g}_{im}\bar{g}_{jn}\varepsilon_{mnk} {\cal L}_k,
\qquad
\{{\cal N}_i, {\cal L}_j\}= \bar{g}_{im}\bar{g}_{jn}\varepsilon_{mnk} {\cal N}_k,
\end{gather*}
where $i,j,k = 1,2,3$. There are two Casimir invariants, the f\/irst of which vanishes in realiza\-tion~(\ref{classical1}):
\begin{gather}
\label{CASIMIR-01}
{\cal C}_1 =  \textbf{L}  \cdot \textbf{N} = \textbf{N}  \cdot \textbf{L} = \bar{g}_{ik} {\cal N}_i {\cal L}_k = {\cal N}_1 {\cal L}_1
- {\cal N}_2 {\cal L}_2 - {\cal N}_3 {\cal L}_3 = 0,
\end{gather}
and the second one is
\begin{gather}
\label{CASIMIR-02}
{\cal C}_2  =  N^2 + L^2,
\end{gather}
where
\begin{gather}
N^2 = \textbf{N}  \cdot \textbf{N} = \bar{g}_{ik} {\cal N}_i {\cal N}_k =  {\cal N}_1^2 - {\cal N}_2^2 - {\cal N}_3^2,
\nonumber\\
L^2 = \textbf{L}  \cdot \textbf{L} = \bar{g}_{ik} {\cal L}_i {\cal L}_k = {\cal L}_1^2 - {\cal L}_2^2 -{\cal L}_3^2.\label{0LB:1}
\end{gather}
The next step is computing the relationship between the ambient momenta and the geodesic polar one.
Taking into account that four-dimensional canonical momentum $p_\mu$ ($\mu = 0,1,2,3$)
\begin{gather*}
p_\mu = \frac{\partial {\cal L}}{\partial  \dot{z}^\mu} = G_{\mu \nu} \dot{z}^\nu ,
\qquad
{\cal L}  = \frac{1}{2} {G}_{\mu \nu} \dot{z}^\mu \dot{z}^\nu,
\end{gather*}
where ${\cal L}$ is a kinetic energy in the ambient space {\bf R}$_{2,2}$, we obtain that
\begin{gather*}
R \cdot p_0 = - R \cdot\frac{\partial z_0}{\partial t} = - \sinh r \, p_r,
\\
R \cdot p_1 = - R \cdot\frac{\partial z_1}{\partial t} = - \cosh r \sinh \tau \, p_r + \frac{\cosh \tau}{\sinh r} \, p_\tau,
\\
R \cdot p_2 =   R \cdot \frac{\partial z_2}{\partial t} =   \cosh r \cosh \tau \cos\varphi \, p_r -  \frac{\sinh \tau \cos\varphi}{\sinh r} \, p_\tau
- \frac{\sin\varphi}{\sinh r \cosh\tau} p_\varphi ,
\\
R \cdot p_3 =  R \cdot \frac{\partial z_3}{\partial t} =   \cosh r \cosh \tau \sin\varphi \, p_r -  \frac{\sinh \tau \sin\varphi}{\sinh r} \, p_\tau
+ \frac{\cos\varphi}{\sinh r \cosh\tau} p_\varphi.
\end{gather*}
Then the generators (\ref{classical1}) in geodesic pseudo-spherical coordinates and momenta are given by the formulas
\begin{gather}
{\cal N}_1  = - \sinh \tau  \, p_r + \cosh\tau \coth r \, p_\tau,
\nonumber\\
{\cal N}_2   =  - \cosh \tau \cos \varphi \, p_r + \coth r \sinh \tau \cos\varphi \, p_\tau + \frac{\coth r \sin\varphi}{\cosh\tau} \, p_\varphi,
\nonumber\\
{\cal N}_3   =  - \cosh \tau \sin \varphi \, p_r + \coth r \sinh \tau \sin\varphi \, p_\tau - \frac{\coth r \cos\varphi}{\cosh\tau} \, p_\varphi,
\nonumber\\
{\cal L}_3   =   - \cos\varphi \, p_\tau +  \frac{\sin\varphi}{\coth \tau} \, p_\varphi,
\qquad
{\cal L}_2   =   - \sin\varphi \, p_\tau -  \frac{\cos\varphi}{\coth \tau} \, p_\varphi,
\qquad
{\cal L}_1   =  p_\varphi.\label{ANGULAR-03}
\end{gather}
Using now equations (\ref{CASIMIR-02}), (\ref{0LB:1}) and (\ref{ANGULAR-03}) it is easy to see the second Casimir operator~${\cal C}_2$ is related
with the free Hamiltonian (\ref{LAGRANGE-04}) by ${\cal C}_1 = - 2 R^2 {\cal H}_{\rm free}$. Thus all the quantities (\ref{ANGULAR-03}) Poisson commute
with free Hamiltonian (\ref{LAGRANGE-04}) and are constants of the motion. From the seven integrals of the motion $\{{\cal H}_{\rm free}, {\cal N}_i, {\cal L}_i\}$ only f\/ive are functionally independent, because of the relation~(\ref{CASIMIR-02}) and constraint~(\ref{CASIMIR-01}).
Hence the geodesic motion  with the Hamiltonian~(\ref{LAGRANGE-04}) turns out to be a {\it maximally superintegrable system}.

Let us now consider the spherically symmetric model, namely the Hamiltonian  ${\cal H} = {\cal H}_{\rm free} + {\cal V}(r)$, where ${\cal H}_{\rm free}$ is
given by equation~(\ref{LAGRANGE-04}) and ${\cal V}(r)$ is a potential function.
It is obvious that the Hamilton--Jacobi equation ${\cal H} = {\cal E}$ for any central potential admit separation of variables in the pseudo-spherical
system of coordinates (\ref{coor:1}) (and (\ref{PSEUDO:2}))\footnote{Beside of the pseudo-spherical system of coordinates~(\ref{coor:1}) the Hamilton--Jacobi equation ${\cal H}_{\rm free} = {\cal E}$
and free Schr\"odinger equation on $H_2^2$ hyperboloid allow the separation of variables additionally in 70th orthogonal systems of coordinates
(see for details~\cite{KALMIL1}).}.
The pseudo-spherical system of coordinates corresponds to the subgroup chains  ${\rm SO}(2,2) \supset {\rm SO}(2,1)  \supset {\rm SO}(2)$.
Thus, the central symmetry of Hamiltonian~${\cal H}$ implies the conservation low of the vector $\textbf{L} = ({\cal L}_1, {\cal L}_2, {\cal L}_3)$
with the scalar product~(\ref{CASIMIR-01}), which we can interpreted as Lorenzian ``angular momentum''.
In particular the f\/irst component of  angular momentum ${\cal L}_1 = p_\varphi$ and Casimir invariant of algebra ${\mathfrak{so}}(2,1)$:
\begin{gather}
\label{ANGULAR-02}
L^2  = {\cal L}_1^2 - {\cal L}_2^2 - {\cal L}_3^2  = - \left(p_\tau^2 -  \frac{p_\varphi^2}{\cosh^2\tau}\right),
\end{gather}
together with the Hamiltonian ${\cal H}$:
\begin{gather*}
{\cal H}  = \frac{1}{2 R^2}\left\{p_r^2 + \frac{L^2}{\sinh^2 r}\right\} +  {\cal V}(r),
\end{gather*}
form the mutually Poisson-involutive system of constants of motion.
As it follows from the equation (\ref{ANGULAR-02}): ${p^2_\varphi}/{\cosh^2\tau} - L^2  \geq 0$, the quantity $L^2$,
in contrast to the motion in Euclidean space (or spheres and two-sheeted hyperboloids), can take not only the positive or zero
but also the negative value. Another dif\/ference is that at the f\/ixed values of~$L^2$: $p^2_\varphi \geq L^2$.
The existence of an additional independent constant of motion ${\cal L}_2$ (${\cal L}_3$ then not independent)
means that the problem is at least once degenerate and the trajectories placed on the two-dimensional surface.
For the case of positive $L^2$ putting $\tau = 0$, or $L^2 = p^2_\varphi$, we obtain that the motion takes place
on the two-dimensional subspace, namely two-sheeted hyperboloid $z_0^2  - z_2^2 - z_3^2 = R^2$, while for negative~$L^2$, we may put $\varphi = 0$ or $p^2_\varphi = 0$, and restricted to the one-sheeted hyperboloid
$z_0^2  + z_1^2 - z_2^2 = R^2$.

In the case of $|z_0| < R$ the formulas for ${\mathfrak{so}}(2,2)$ generators~(\ref{ANGULAR-03}) are changed accordingly to the transformation (\ref{TRANSO-01}).
We have
\begin{gather*}
L^2  = - \left(p_\mu^2 +  \frac{p_\varphi^2}{\sinh^2\mu}\right).
\end{gather*}
Hence by virtue of above relation, the $L^2$ takes only negative value. Without the loss of generality we can put
$\varphi=0$ or $p^2_\varphi = 0$ and the motion on $H_2^2$ again restricted to the one-sheeted hyperboloid
$z_0^2  + z_1^2 - z_2^2 = R^2$.

\section{Harmonic oscillator potential}

Let us now concentrate on the spherically symmetric model, namely harmonic oscillator system.
In the article~\cite{PETPOG3} we have extended the Euclidean isotropic harmonic oscillator potential with the frequency
$\omega$ to our space~$H_2^2$, which is given by
\begin{gather*}
V^{\rm osc} = \frac{\omega^2 R^2}{2}\left(
\frac{z_2^2+z_3^2-z_1^2}{z_0^2}\right) =
\begin{cases}  \dfrac{\omega^2 R^2}{2} \tanh^2 r, & |z_0| \geq R,
\vspace{1mm}\\
  - \dfrac{\omega^2 R^2}{2} \tan^2 \chi, &  |z_0| \leq R.
\end{cases}
\end{gather*}
Respectively the Hamiltonian may be expressed as follow
\begin{gather}
\label{eq:1}
{\cal H}^{\rm osc} = \frac{1}{2R^2}\left(p_r^2 + \frac{L^2}{\sinh^2r}\right) + \frac{\omega^2R^2}{2} \tanh^2r
\end{gather}
for $|z_0| \geq R$, and
\begin{gather}
\label{eq:01}
{\cal H}^{\rm osc} = - \frac{1}{2R^2}\Big(p_\chi^2 + \frac{L^2}{\sin^2\chi}\Big) - \frac{\omega^2R^2}{2} \tan^2 \chi
\end{gather}
for $|z_0| \leq R$.

The Hamiltonian of the harmonic oscillator system, besides the angular momentum~$\textbf{L}$
has additional integrals of motion quadratic in the momenta, which are associated with the generators
$({\cal N}_1, {\cal N}_2, {\cal N}_3)$, the so called Demkov--Fradkin tensor~\cite{DEMKOV,FRADK}:
\begin{gather*}
{\cal D}_{ik} = \frac{1}{R^2} {\cal N}_i {\cal N}_k  + \omega^2 R^2 \frac{z_i z_k}{z_0^2},
\qquad
{\cal D}_{ik} = {\cal D}_{ki},
\qquad
i,k = 1,2,3.
\end{gather*}
The components of ${\cal D}_{ik}$ tensor Poisson commute with Hamiltonian of harmonic oscillator~(\ref{eq:1}) and~(\ref{eq:01}), but not necessarily with each other. In the pseudo-spherical coordinates
the diagonal components of this tensor has the form
\begin{gather*}
{\cal D}_{11} = \frac{{\cal N}_1^2}{R^2} +  \omega^2 R^2  \sinh^2\tau \tanh^2 r,
\qquad
{\cal D}_{22} = \frac{{\cal N}_2^2}{R^2} +  \omega^2 R^2 \cosh^2\tau\cos^2\varphi\tanh^2 r,
\\
{\cal D}_{33} = \frac{{\cal N}_2^2}{R^2} +  \omega^2 R^2 \cosh^2\tau\sin^2\varphi\tanh^2 r,
\end{gather*}
so the harmonic oscillator Hamiltonian is given by
\begin{gather}
\label{DEMKOV-02}
{\cal H}^{\rm osc} =  - {\cal D}_{11} + {\cal D}_{22} + {\cal D}_{33} - \frac{L^2}{2 R^2}.
\end{gather}
In addition to this, the Demkov--Fradkin tensor has the algebraic properties
\begin{gather}
\label{DEMKOV-03}
\sum_{i}{\cal L}_i {\cal D}_{ik} = \sum_{i} {\cal D}_{k i} {\cal L}_i = 0,
\qquad
k = 1,2,3.
\end{gather}
It is clear that the ten integrals of motion $\{{\cal H}, {\cal L}_i, {\cal D}_{ik}\}$
cannot be functionally independent because of the relations~(\ref{DEMKOV-02}) and~(\ref{DEMKOV-03}),
and that
\begin{gather*}
\{ {\cal L}_1 {\cal D}_{11} \} = \{ {\cal L}_2 {\cal D}_{22} \} =  \{ {\cal L}_3 {\cal D}_{33} \} = 0.
\end{gather*}
Only f\/ive integrals of motion, which we can choose as $\{{\cal H}, L^2, {\cal L}_1, {\cal L}_2, {\cal D}_{33}\}$,
are functionally independent. Thus ${\cal H}^{\rm osc}$ is a maximally superintegrable Hamiltonian.
The components of angular momentum and Demkov--Fradkin tensor forms the quadratic algebra. The nonvanishing
Poisson brackets have been presented in Appendix~\ref{appendixA}.

In the contraction limit $R\!\to \!\infty$ the $H_2^2$ hyperbolic space turns into the Minkowski space~{\bf M}$^{2+1}$.
Let us pass to Beltrami coordinates
\begin{gather}
\label{OSC:001}
x_i=R\frac{z_i}{z_0} = R \frac{z_i}{\sqrt{R^2+z_2^2+z_3^2-z_1^2}},
\qquad i = 1,2,3.
\end{gather}
Then, at the limit $R\to \infty$ we have that
\begin{gather*}
\lim_{R\to\infty} V^{\rm osc}(r) = \frac{\omega^2}{2} \big({-}x_1^2+x_2^2+x_3^2\big),
\end{gather*}
which can be interpreted as a harmonic oscillator potential on the ${\bf M}^{2+1}$ Minkowski space $(x_1, x_2, x_3)$.

\section{Integration of the Hamilton--Jacobi equation}

The Hamilton--Jacobi equation, associated with the Hamiltonian~(\ref{eq:1}), is obtained
after the substitution $p_{\mu_i} \to \partial S/ \partial\mu_i$,
where $\mu_i = (r, \tau, \varphi)$. Therefore we get
\begin{gather*}
{\cal H} = \frac{1}{2R^2}\left\{\left(\frac{\partial S}{\partial r}\right)^2 - \frac{1}{\sinh^2r}
\left(\frac{\partial S}{\partial \tau}\right)^2 +
\frac{1}{\sinh^2r\cosh^2\tau}\left(\frac{\partial S}{\partial \varphi}\right)^2\right\}
+ \frac{\omega^2R^2}{2} \tanh^2r = E.
\end{gather*}
This equation is completely separable, and the coordinate $\varphi$ is cyclic. We look the solution for the classical action
$S(r, \tau, \varphi)$ in form
\begin{gather*}
S(r, \tau, \varphi) = p_\varphi \varphi + S_1(r) + S_2(\tau) - E t,
\end{gather*}
and obtain
\begin{gather}
\label{H-J3}
\left(\frac{\partial S_2}{\partial \tau}\right)^2 - \frac{p^2_\varphi}{\cosh^2\tau}
 =  - L^2,
\\
\label{H-J3-1}
\frac{1}{2R^2}\left(\frac{\partial S_1}{\partial r}\right)^2 +
\frac{\omega^2R^2}{2} \tanh^2r + \frac{L^2}{2R^2\sinh^2r} = E.
\end{gather}
The ``quasi-radial'' equation (\ref{H-J3-1}) describes the motion in f\/ield of ef\/fective potential
\begin{gather}
\label{H-EFF-0}
U_{\text{ef\/f}}(r) =  \frac{\omega^2R^2}{2} \tanh^2r +  \frac{L^2}{2R^2\sinh^2r}.
\end{gather}
At the large $r \sim \infty$ the ef\/fective potential $U_{\text{ef\/f}}(r)$ tends to a constant value equal to
$\omega^2R^2/2$, whereas the behavior at the point $r=0$ is determined by the angular momentum~$L^2$.

\begin{figure}[t]
\centering
\includegraphics[width=3.5in]{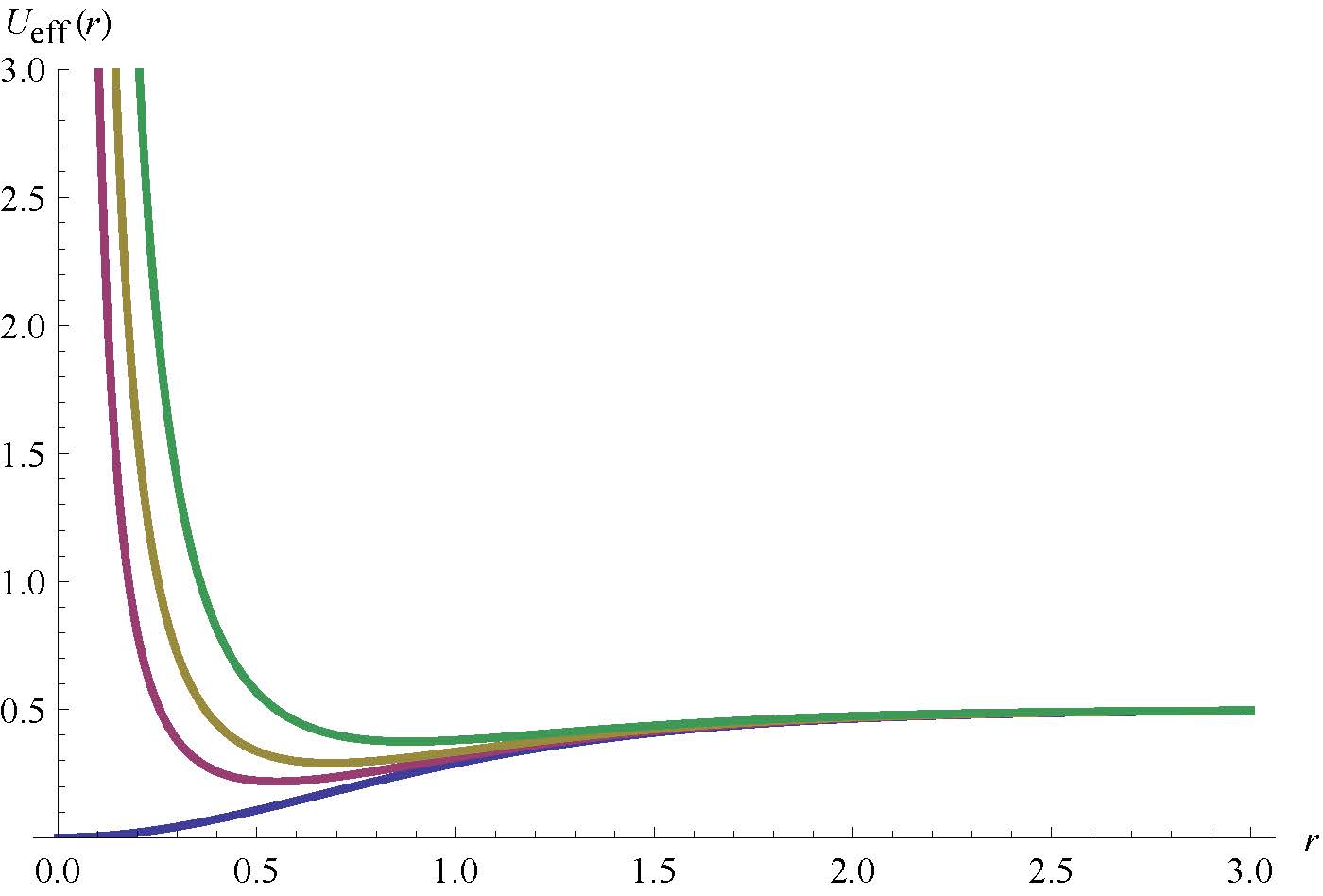}
\caption{Ef\/fective potential $U_{\text{ef\/f}}(r)$ in case of $0 \leq L^2 < \omega^2 R^4$ for value of  $L^2 = 0, 1/16, 1/8, 1/4$;  $\omega = R = 1$.}
\label{Fig1}
\end{figure}

\begin{figure}[t!]
\centering
\includegraphics[width=3.5in]{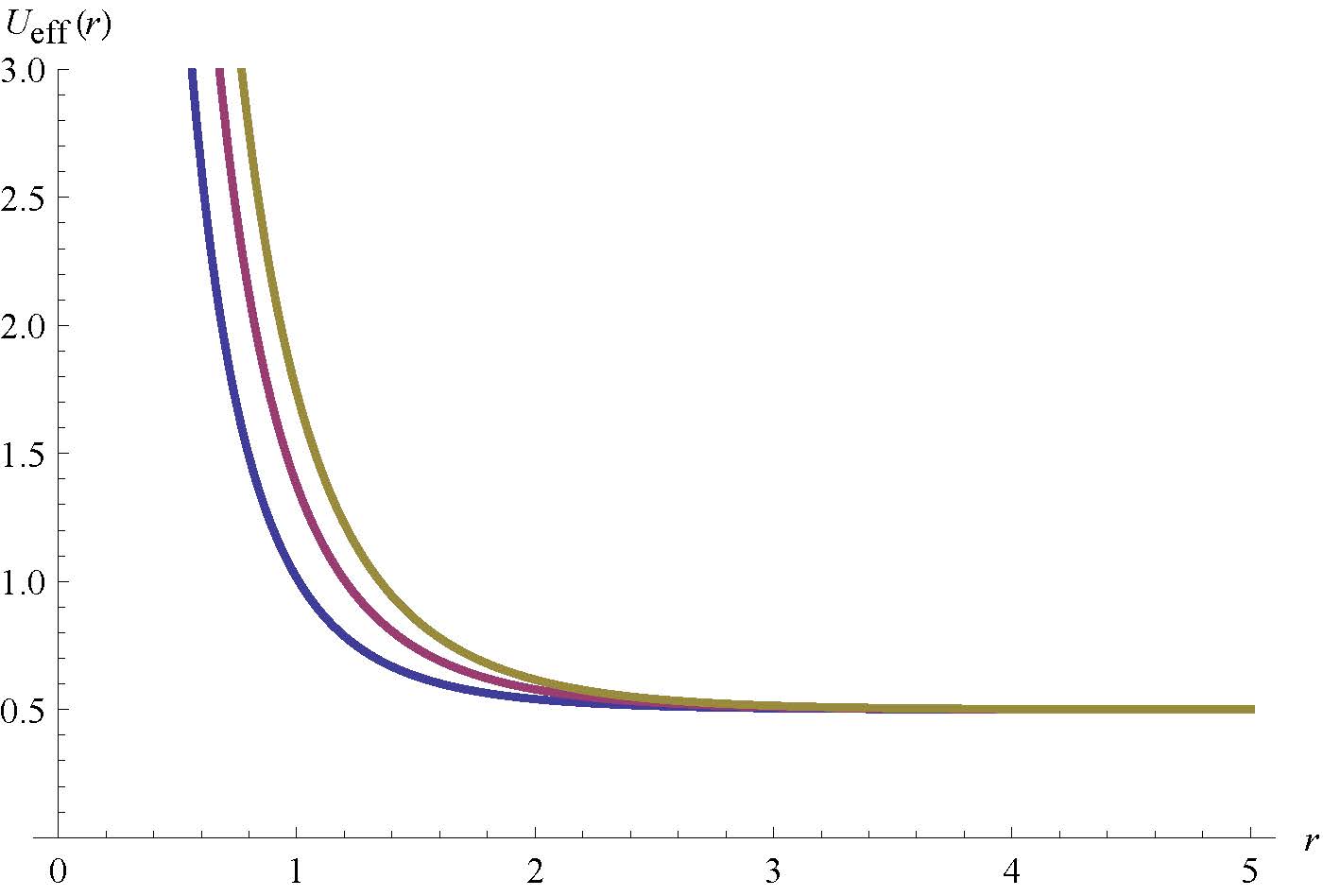}
\caption{Ef\/fective potential $U_{\text{ef\/f}}(r)$ in case of $L^2 \geq \omega^2 R^4$ for value of $L^2 =2, 3, 4$;  $\omega = R=1$.}
\label{Fig2}
\end{figure}

\begin{figure}[t]
\centering
\includegraphics[width=3.5in]{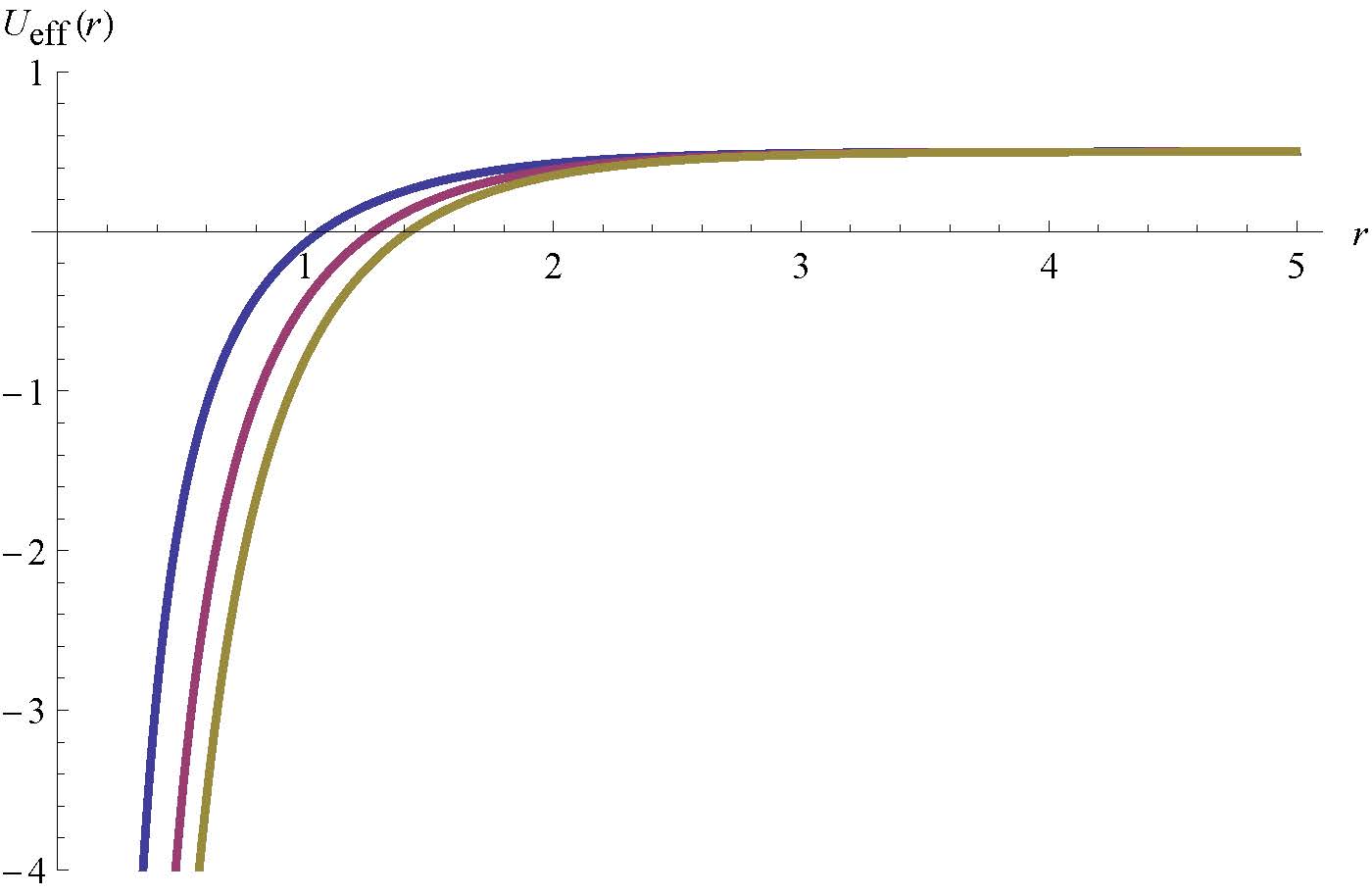}
\caption{Ef\/fective potential $U_{\text{ef\/f}}(r)$ in case of $L^2 < 0$ for value of $L^2 = -1, -2, -3$; $\omega=R=1$.}\label{Fig3}
\end{figure}

In case $0 \leq L^2 < \omega^2 R^4$ potential (\ref{H-EFF-0}) has a minimum at $r_0 = \tanh^{-1}\sqrt[4]{L^2/\omega^2 R^4}$
(see Fig.~\ref{Fig1}), and at this point
\begin{eqnarray}
\label{H-EFF-1}
0 \leq U_{\text{ef\/f}} (r_0) =  {\omega}{\sqrt{L^2}} - \frac{L^2}{2 R^2} < \frac{\omega^2R^2}{2},
\end{eqnarray}
where equality is possible only in case of $L^2 = 0$. For $L^2 \geq \omega^2 R^4$
the potential $U_{\text{ef\/f}}(r)$ is repulsive on the whole semi-axis $r \in [0, \infty)$ (see Fig.~\ref{Fig2}).
In the case of negative $L^2$ the ef\/fective potential~(\ref{H-EFF-1}) is attractive and has
a singularity for a small $r$ as $\sim r^{-2}$ (see Fig.~\ref{Fig3}).

For the region $|z_0| < R$  the dif\/ferential equations~(\ref{H-J3}) and~(\ref{H-J3-1}) are transformed to
the following ones
\begin{gather*}
\left(\frac{\partial S_2}{\partial \mu}\right)^2 + \frac{p^2_\varphi}{\sinh^2\mu}
 =  - L^2,
\\
\frac{1}{2R^2}\left(\frac{\partial S_1}{\partial \chi}\right)^2 +
\frac{\omega^2R^2}{2} \tan^2\chi + \frac{L^2}{2R^2\sin^2\chi} = - E.
\end{gather*}
The f\/irst equation admits only negative value of~$L^2$. Therefore we take into account the motion inside the region
$|z_0| < R$ when investigate we the case of negative value of~$L^2$.

Integrating now equations (\ref{H-J3}) and (\ref{H-J3-1}) we get
\begin{gather}
\label{H-J4-1}
S_1(r) =  \int\sqrt{2R^2E- \omega^2 R^4\tanh^2r-\frac{L^2}{\sinh^2r}}dr,
\\
\label{H-J4-2}
S_2(\tau)  =  \int\sqrt{-  L^2 +\frac{p^2_\varphi}{\cosh^2\tau}}d\tau.
\end{gather}
Since we are interested only the trajectories we will follow the usual procedures~\cite{LL} and consider the equations
\begin{gather}
\label{H-J5}
\frac{\partial S}{\partial E} = \frac{\partial S_1}{\partial E} - t = - t_0,
\qquad
\frac{\partial S}{\partial L^2}= \frac{\partial S_1}{\partial L^2} +
\frac{\partial S_2}{\partial L^2} = \beta,
\qquad
\frac{\partial S}{\partial
p_\varphi} = \varphi +  \frac{\partial S_2}{\partial
p_\varphi}= \varphi_0,
\end{gather}
where $t_0$, $\varphi_0$ and $\beta$ are the constants.

\subsection{Integration of quasi-radial part}

From equations (\ref{H-J4-1}) and (\ref{H-J5}) we get that
\begin{gather}
\label{H-J6}
t-t_0 =  \frac{1}{\omega}   \int\frac{\tanh r dr}{\sqrt{- \tanh^4r + 2\left(E /\omega^2R^2 + L^2/2 \omega^2 R^4\right)
\tanh^2 r - L^2/\omega^2 R^4}}.
\end{gather}
Below we consider separately  all four cases: $0 < L^2 < \omega^2 R^4$, $L^2 \geq \omega^2 R^4$,  $L^2 < 0$ and $L^2=0$.

{\bf 1.} The case  $0 < L^2 < \omega^2 R^4$.
For the roots in the radical expression of denominator in (\ref{H-J6}) we have
\begin{gather}
\label{H-J6a}
X_{1,2}= \frac{(2R^2 E + L^2) \pm \sqrt{(2 R^2 E + L^2)^2 -  4 L^2 \omega^2 R^4}}{2 \omega^2 R^4},
\end{gather}
where $X=\tanh^2r \in [0,1]$. It's obvious that the radicand in equation (\ref{H-J6a}) is positive for any values
of energy $E > E_{\min} = U_{\text{ef\/f}} (r_0)$ and equal zero for $E = E_{\min}$. Thus the roots $X_{1,2}$ $(X_{1} \leq X_{2})$
are positive. It is easy to see that for $E_{\min} \leq E < \omega^2R^2/2$ both roots satisfy the inequality condition
$0 < X_1 <  X_2 < 1$. At $E \geq \omega^2R^2/2$: $0 < X_1 <  1 \leq X_2$ and equality $X_2 = 1$ is possible only
for $E = \omega^2R^2/2$. The bounded motion exists exclusively for $E_{\min} \leq E < \omega^2R^2/2$.
Below we will consider separately all possible cases,
namely:  $E_{\min} < E < \omega^2R^2/2$, $E = E_{\min}$, $E > \omega^2R^2/2$ and $E = \omega^2R^2/2$.

{\bf A.} Performing the integration in formula~(\ref{H-J6}) we get for  $E_{\min}< E < \omega^2R^2/2$
\begin{gather*}
2\omega^2R^2 \sinh^2 r
 =
(1-2E/\omega^2R^2)^{-1} \Bigl\{\big(2 E - L^2/R^2\big)
\\
\hphantom{2\omega^2R^2 \sinh^2 r=}{} +
\sqrt{\big(2 E + L^2/R^2\big)^2- 4L^2 \omega^2}   \sin\big[2 \omega \sqrt{1- 2 E/\omega^2R^2}(t-t_0)\big]\Bigr\}.
\end{gather*}
Thus the motion is bounded and periodic.  The period is given by
\begin{gather}
\label{H-J7-02}
T (R) =  \frac{\pi}{\omega} \frac{1}{\sqrt{1 - 2E/\omega^2 R^2}}.
\end{gather}
The total frequency $\omega_0 = \omega\sqrt{1 - 2E/\omega^2 R^2}$ and unlike the motion in Euclidean space,
depends on the energy of particle $E$ and curvature of the space $ \kappa = - 1/R^2$ as a parameter,
but it is constant for each of the orbits at a f\/ixed value of the energy\footnote{The Euclidean harmonic oscillator
is a classical example of an {\it isochronous} system~\cite{RANADA2}.
The period of motion of Euclidean oscillator depends only from frequency and is the same for all orbit.}.
This property is common to all
closed orbits of superintegrable systems on the spaces with constant curvature.
The contraction limit~$R\to \infty$ give us the correct Euclidean period: $T(R)_{R\to \infty} =  \frac{\pi}{\omega}$.
The period of motion on~$H^2_2$ always larger than in Euclidean space by the factor: $1/\sqrt{1 - 2E/\omega^2 R^2}$
and tends to inf\/inity at the limit $E \to \omega^2 R^2/2$, that is the closed orbits changes to the inf\/inite open ones.

{\bf B.} In the case of minimum energy: $E = E_{\min} = U_{\text{ef\/f}}(r_0)$ or $E_{\min} = {\omega}{\sqrt{L^2}} - {L^2}/{2 R^2}$
the integral in~(\ref{H-J6}) is not def\/ined and we must solve directly the equation~(\ref{H-J3-1}).
From equation~(\ref{H-J3-1}) we obtain
\begin{gather*}
\left(\frac{\partial S_1}{\partial r}\right)^2 =
-   \left(\sqrt{L^2}\coth r - \sqrt{\omega^2 R^4}\tanh r\right)^2
\geq 0,
\end{gather*}
or ${\partial S_1}/{\partial r}=0$ and $\tanh^2 r = \sqrt{L^2/\omega^2 R^4}$. Therefore
\begin{gather}
\label{H-J10}
r = \tanh^{-1} \left(\sqrt{1-\sqrt{1-\frac{2 E}{\omega^2R^2}}}\right),
\end{gather}
i.e., the trajectories are circles. Here from two values of $\sqrt{L^2}$ allowed by equation $E=U_{\text{ef\/f}}(r_0)$,
we choose the smaller one $\sqrt{L^2}= \omega R^2 \left(1 - \sqrt{1 - 2E/\omega^2 R^2}\right)$
because it satisf\/ies the condition $0 < L^2 < \omega^2 R^4 $. In case of  contraction limit $R \to \infty$
we obtain $E=E_{\min} = {\omega}{\sqrt{L^2}}$ and $r = \sqrt{E}/\omega$.

{\bf C.} In case of $E>\omega^2R^2/2$ after integration in~(\ref{H-J6})  we have
\begin{gather}
2\omega^2R^2 \sinh^2 r
=
\big(2E/\omega^2R^2 - 1\big)^{-1} \Bigl\{\big(L^2/R^2 - 2 E\big)
\nonumber
\\
\hphantom{2\omega^2R^2 \sinh^2 r=}{}
+
\sqrt{\big(2 E + L^2/R^2\big)^2- 4L^2 \omega^2}   \cosh\big[2 \omega \sqrt{2 E/\omega^2R^2 - 1}(t_0-t)\big]\Bigr\},\label{H-J7-003}
\end{gather}
i.e.,  the motion is not bounded.

{\bf D.} For the limiting case of $E=\omega^2R^2/2$ the roots of denominator are $X_1 = L^2/\omega^2 R^4$, $X_2= 1$, thus
$L^2/\omega^2 R^4 < \tanh^2 r < 1$ and motion is not bounded because of ${\tanh^{-1}}(L^2/\omega^2 R^4) < r < \infty$.
The integration in~(\ref{H-J6}) yield
\begin{gather}
\label{H-J10-1}
\cosh^2 r = \big(1-L^2/\omega^2R^4\big)^{-1} + \omega^2 \big(1-L^2/\omega^2R^4\big)   (t-t_0)^2.
\end{gather}

{\bf 2.} Let us consider now the case of $L^2 \geq \omega^2 R^4$ (see Fig.~\ref{Fig2}).
From equation~(\ref{H-J6}) we get that the only possible value for energy is $E > \omega^2 R^2/2$
and the roots satisfy the inequality $0 < X_1 < 1 < X_2$. Thereby, the equation of motion
is determined by the formula~(\ref{H-J7-003}). The motion of particle is limited only by the point
$r_{\min} = \tanh^{-1}\sqrt{X_1}$, i.e., it has the ability to go to inf\/inity.

{\bf 3.} Let us consider f\/inally the case of $L^2 \leq 0$. From the equation (\ref{H-J6}) we have
that the roots of denominator are
\begin{gather*}
X_{1,2} = \frac{\left(2 E R^2 - |L^2|\right) \pm \sqrt{\left(2 E R^2 -  |L^2|\right)^2 +  4|L^2| \omega^2 R^4}}{2 \omega^2 R^4},
\end{gather*}
where again $X = \tanh^2r \in [0,1]$.
It can be seen that $X_1 < 0 < X_2$ is independent of the value of $A$ and energy $E$.
For the region $E \geq \omega^2R^2/2$ one of the roots is $X_2>1$, so the radicand is positive for
any values of variable $r$, including the point $r=0$: $r \in [0, \infty)$.
The same situation develops for region $E< \omega^2R^2/2$,  where $r \in [0,  {\tanh}^{-1}\sqrt{X_2}]$.
Therefore in case of negative $A$ the particle can penetrate from the region $z_0 \geq R$ to
$0 \leq z_0 \leq R$.

Performing the integration in formula (\ref{H-J6}), we have for $E<\omega^2R^2/2$
\begin{gather}
\sinh^2 r  =  \frac{2R^2 E + |L^2|}{2R^2(\omega^2 R^2 - 2E)}
\nonumber\\
\hphantom{\sinh^2 r  =}{} +  \frac{\sqrt{(2R^2 E - |L^2|)^2 + 4|L^2|\omega^2 R^4}}{2R^2(\omega^2 R^2 - 2E)}
\sin\left[2\omega\sqrt{1- 2E/\omega^2R^2}(t-t_0)\right],\label{H-J13-0}
\end{gather}
while for $E>\omega^2R^2/2$
\begin{gather}
\sinh^2 r
=  \frac{2R^2 E + |L^2|}{2R^2(\omega^2 R^2 - 2E)}
\nonumber\\
\hphantom{\sinh^2 r=}{}
+ \frac{\sqrt{(2R^2 E - |L^2|)^2 + 4|L^2|\omega^2 R^4}}{2R^2(2E - \omega^2 R^2)}
\cosh\left[2\omega\sqrt{2E/\omega^2R^2-1}(t-t_0)\right].\label{H-J7-3}
\end{gather}
From the formula (\ref{H-J13-0}) it follows that the motion at $E<\omega^2R^2/2$
is bounded and periodic with period~(\ref{H-J7-02}).
Below we will construct the bounded trajectories lying on the whole hyperboloid, namely
not only in the region $|z_0| \geq R$, but also $|z_0| \leq R$. In case when $E = \omega^2R^2$ the
integration in~(\ref{H-J6}) leads, up to a~transformation $L^2 \to - |L^2|$, to a~result
similar to the formula~(\ref{H-J10-1}).

In the limiting case of $L^2=0$ the formulas (\ref{H-J13-0}), (\ref{H-J7-3}) and~(\ref{H-J10-1})
are simplif\/ied. For  $0 < E < \omega^2R^2/2$ we get
\begin{gather*}
\sinh^2 r  =
\frac{2E/\omega^2R^2}{1-2E/\omega^2R^2}
\cos^2\left(\omega \sqrt{1- 2 E/\omega^2R^2}(t-t_0) - \frac{\pi}{4}\right),
\end{gather*}
while in case of $E > \omega^2R^2/2$
\begin{gather*}
\sinh r  =
\sqrt{\frac{2E/\omega^2R^2}{2E/\omega^2R^2 - 1}}
\sinh\left(\omega \sqrt{2 E/\omega^2R^2 - 1}(t_0-t)\right).
\end{gather*}
Finally for $E = \omega^2R^2/2$ we obtain  $\sinh r  = \omega (t-t_0)$.

\subsection{Integration of the angular parts}

{\bf 1.}   Let us f\/irst consider the case when $L^2 > 0$.  From (\ref{H-J4-1}) and~(\ref{H-J4-2})
we obtain
\begin{gather}
\label{H-J14}
\frac{\partial S_1}{\partial L^2}
= - \frac{1}{2}\int\frac{dr}{\sinh^2r\sqrt{2R^2E - \omega^2 R^4\tanh^2r - L^2/\sinh^2r}},
\\
\label{H-J15}
\frac{\partial S_2}{\partial L^2}
= - \frac12\int\frac{d\tau}{\sqrt{- L^2 + p_\varphi^2/\cosh^2\tau}}.
\end{gather}
The integrals can be easily calculated to give~\cite{RUJIK}
\begin{gather*}
\frac{\partial S_2}{\partial L^2} = - \frac{1}{\sqrt{4L^2}}
 \arcsin \left[\frac{\sinh\tau}{\sqrt{p_{\varphi}^2/L^2-1}}\right],
\\
\frac{\partial S_1}{\partial L^2}
=  \frac{1}{4\sqrt{A}}
 \arcsin \left[\frac{2L^2 \coth^2 r - (2E R^2 + L^2)}{\sqrt{(2E R^2 + L^2)^2 - 4L^2 \omega^2 R^4}}\right].
\end{gather*}
Here we require
\begin{gather*}
\label{H-J18}
- \sqrt{p_{\varphi}^2/L^2-1} < \sinh \tau <  \sqrt{p_{\varphi}^2/L^2-1},
\end{gather*}
and
\begin{gather}
\label{H-J19}
\big|2L^2 \coth^2 r - \big(2ER^2 + L^2\big)\big| < \sqrt{\big(2E R^2 + L^2\big)^2 - 4L^2 \omega^2 R^4}.
\end{gather}
The condition (\ref{H-J19}) is equivalent to $z_1 < \coth r < z_2$, where $z_{1,2}$ are the roots
of denominator in integral (\ref{H-J14}):
\begin{gather*}
z_{1,2}= \frac{\big(2E R^2 + L^2\big) \pm \sqrt{\big(2E R^2 +  L^2\big)^2 - 4 L^2 \omega^2 R^4}}{2L^2},
\qquad
E \geq E_{\min} = {\omega}{\sqrt{L^2}} - {L^2}/{2 R^2}.
\end{gather*}
The f\/inal condition $z_2 > 1$ implies that $L^2 > \omega^2 R^4$ and $E > \omega^2 R^2 /2$ or $0 < L^2 < \omega^2 R^4$ and $E > E_{\min}$.

Therefore for ${\partial S}/{\partial L^2}$ we have
\begin{gather}
\frac{\partial S}{\partial L^2}
=
\frac{1}{4\sqrt{L^2}}\left\{\arcsin \left[\frac{2L^2 \coth^2 r - \big(2E R^2 + L^2\big)}{\sqrt{\big(2E R^2 + L^2\big)^2 - 4L^2 \omega^2 R^4}}\right]
\right.\nonumber\\
\left.\hphantom{\frac{\partial S}{\partial L^2}=}{}
-
2 \arcsin \left[\frac{\sinh\tau}{\sqrt{p_{\varphi}^2/L^2-1}}\right]
\right\} = \beta.\label{H-J20}
\end{gather}
Next, from (\ref{H-J4-2}) and (\ref{H-J5}) we obtain
\begin{gather}
\label{H-J22}
\frac{\partial S}{\partial p_\varphi}  = \varphi+\int\frac{p_\varphi
d\tau}{\cosh^2\tau\sqrt{-L^2 + p_\varphi^2/\cosh^2\tau}} =
\varphi+\arcsin\frac{\tanh\tau}{\sqrt{1 - L^2/p_\varphi^2}}
= \varphi_0,
\end{gather}
and hence
\begin{gather}
\label{H-J23}
\tanh\tau = \sqrt{1 - L^2/p_\varphi^2}   \sin(\varphi_0-\varphi).
\end{gather}

{\bf 2.}  Let us consider the integration in formulas (\ref{H-J14}), (\ref{H-J15}) and
(\ref{H-J22}) in the case $L^2 \leq 0$. Instead of equation (\ref{H-J20}) we obtain \cite{RUJIK}
\begin{gather}
\frac{\partial S}{\partial L^2}
=
\frac{1}{4\sqrt{|L^2|}}
\left\{
\arccosh \left[\frac{2|L^2| \coth^2 r + \big(2E R^2 - |A|\big)}{\sqrt{\big(2E R^2 - |L^2|\big)^2 + 4|L^2| \omega^2 R^4}}\right]\right.
\nonumber\\
\left.\hphantom{\frac{\partial S}{\partial L^2}=}{}
-
2 \arcsinh \left[\frac{\sinh\tau}{\sqrt{1+p_{\varphi}^2/|L^2|}}\right]\right\}\label{H-J16-0}
= \beta,
\end{gather}
and
\begin{gather}
\label{H-J0-23}
\sin (\varphi_0-\varphi) = \frac{p_\varphi}{\sqrt{p_\varphi^2 + |L^2|}}  \tanh \tau,
\end{gather}
with the restriction for $r$:
\begin{gather*}
\coth^2 r \geq \left(\frac{1}{2} - \frac{E R^2}{|L^2|}\right) + \sqrt{\left(\frac{1}{2} - \frac{E R^2}{|L^2|}\right)^2
+  \frac{\omega^2 R^4}{|L^2|}}.
\end{gather*}
The limiting case of $L^2=0$ could be easily calculated directly from equations (\ref{H-J16-0})
and~(\ref{H-J0-23}). So, we get
\begin{gather}
\label{H-J23-A0}
\left.\frac{\partial S}{\partial L^2}\right|_{L^2=0} =
\frac{\sqrt{2 E \coth^2r-\omega^2R^2}}{ 4 E R} - \frac{\sinh\tau}{2p_\varphi}=\beta,
\qquad
\sinh\tau=\tan(\varphi_0-\varphi)
\end{gather}
with the obvious restriction $\coth^2r \geq \omega^2R^2/2E$.

\section[The trajectories for $L^2 > 0$]{The trajectories for $\boldsymbol{L^2 > 0}$}

From (\ref{H-J20}) and (\ref{H-J23}) we have
\begin{gather}
\label{H-J21}
\coth^2 r =  \left(\frac{ER^2}{L^2}+\frac12\right)+\sqrt{\left(\frac{ER^2}{L^2}+\frac12\right)^2
-\frac{\omega^2 R^4}{L^2}}   \sin\big(2\psi+4\sqrt{L^2}\beta\big),
\end{gather}
where
\begin{gather}
\label{PSI}
\psi = \arcsin{\left(\frac{\sinh\tau}{\sqrt{p_{\varphi}^2/L^2-1}}\right)} =
\arcsin{\left(\frac{1}{\sqrt{1 + L^2/p_{\varphi}^2 \cot^2(\varphi_0-\varphi)}}\right)}.
\end{gather}
Now we can rewrite the equation  (\ref{H-J21}) in form
\begin{gather}
\label{H-J28}
\tanh^2 r =  \frac{1}{\left(\frac{ER^2}{L^2}+\frac12\right)+\sqrt{\left(\frac{ER^2}{L^2}+\frac12\right)^2-\frac{\omega^2 R^4}{L^2}}
\sin\big(2\psi+4\sqrt{L^2}\beta\big)}.
\end{gather}
Thus we see from (\ref{PSI}) that the dependence of angle $\tau$ in the equation of trajectories~(\ref{H-J28}) can be
eliminated. On the other hand from the formula~(\ref{H-J23}) it follows that the motion of particle on the hyperboloid is
restricted to the additional condition
\begin{gather*}
\frac{z_1}{z_3} = \frac{\tanh\tau}{\sin\varphi} = \sqrt{1 - L^2/p_{\varphi}^2}.
\end{gather*}
Therefore, without the loss of generality we can choose $\tau=0$ or $L^2=p_\varphi^2$.
Taking into account that the formula~(\ref{H-J28}) is invariant about transformation
$r \to i\pi - r$ we can conclude that all trajectories of motion, given by
this formula, lie on the upper ($z_0 \geq R$) or lower ($z_0 \leq - R$) sheets of the
two-sheeted hyperboloid: $z_0^2-z_2^2-z_3^2 = R^2$. Obviously they are symmetric with
respect to transformation $z_0 \to - z_0$.

Putting now $L^2=p_\varphi^2$ in (\ref{H-J23}) we obtain that $\psi = (\varphi_0-\varphi)$ and
the formula~(\ref{H-J28}) gain the following form (equation of orbits)
\begin{gather}
\label{H-J29}
\tanh^2 r =  \frac{p}{1 +  \varepsilon (R) \cos 2 \varphi},
\end{gather}
where we use the notations
\begin{gather}
\label{H-J30}
p (R)  = \left(\frac{E R^2}{L^2} +\frac12 \right)^{-1} > 0,
\qquad
\varepsilon (R) = \sqrt{1 -  \frac{4\omega^2 R^4 L^2}{\big(2ER^2+L^2\big)^2}} < 1,
\end{gather}
and choose $\varphi_0 = -2\beta\sqrt{A} + \frac{\pi}{4}$ that the points $\varphi = 0$ will be the nearest
to the center. It is clear that radicand is always positive because of $E>U_{\text{ef\/f}}(r_0)$ for
$0<A< \omega^2 R^4$ and $E > \omega^2R^2/2$ for $A \geq \omega^2 R^4$.

It is well-known that as in the Euclidean plane it is possible to introduce the conic
(section) on the two-dimensional spaces of constant curvature~\cite{chern,DOMBROWSKI,KOZLOV1} (see also the def\/inition of curves
on the two dimensional hyperboloid in~\cite{OLEVSKI}). The conics  on the spaces with constant curvatures are the curves
of the intersection between two-sheeted hyperboloid (or sphere) and second order quadric cone with the origin in the center
of hyperboloid (sphere). Geometrically the conic on the spaces of constant curvature possesses many properties characteristic
of conic section in Euclidean plane, particularly we can speak about the focuses~$F_1$ and~$F_2$ and can
determine the conic as the point set, from which the sum (ellipses) or dif\/ference
(hyperbolas)~$2a$ of distances~$r_1$ and~$r_2$ to two given points (focuses~$F_1$ and~$F_2$)
are constant.

Let us now analysis of the oscillator orbit (\ref{H-J29}).
The formula of trajectories~(\ref{H-J29}) may be written in more convenient form
\begin{gather}
\label{ELL-001}
\frac{1}{\tanh^2r} =  \frac{\cos^2\varphi}{B^2} +  \frac{\sin^2\varphi}{A^2},
\end{gather}
or in term of the Beltrami coordinate (\ref{OSC:001}):
\begin{gather}
\label{ELL-101}
\frac{x_2^2}{B^2} +  \frac{x_3^2}{A^2} = R^2,
\end{gather}
where the constant $A$ and  $B$ are
\begin{gather}
\label{ELL-002}
B^2 = \frac{p (R)}{1 + \varepsilon (R)},
\qquad
A^2 = \frac{p (R)}{1 - \varepsilon (R)},
\qquad
0 < B^2 \leq A^2.
\end{gather}
The orbit equation of the type (\ref{ELL-001}) has been studied in detail in the paper~\cite{RANADA1} (see also \cite{DOMBROWSKI})
at the investigation of two-dimensional harmonic oscillator in the space of constant curvature in polar coordinates.
The curves~(\ref{ELL-001}) are always conic on the hyperbolic plane, but its type depends on the value of $A$ and $B$.
It is obvious that if the value $A^2>1$ and $B^2 >1$, then for any polar angle $\varphi$ it follows that $\tanh r > 1$, and this
case cannot produce any oscillator orbit. In the case of $B^2 < A^2 < 1$ the conic~(\ref{ELL-001}) takes the form of hyperbolic ellipses.
The quantities~$A$ and~$B$ are related to the lengths of the large and small semiaxes~$a$ and~$b$, running the interval $[0, \infty)$,
def\/ined as the values of $r$ at $\varphi = \pi/2$ and $\varphi = 0$. Then the values~$A$,~$B$ can be written in term of hyperbolic tangent
of $a$, $b$: $A^2 = \tanh^2 a$ and $B^2 = \tanh^2 b$ and the equation of orbit~(\ref{ELL-001}) is
\begin{gather}
\label{ELL-003}
\frac{1}{\tanh^2r} =  \frac{\cos^2\varphi}{\tanh^2 b} +  \frac{\sin^2\varphi}{\tanh^2 a}.
\end{gather}
In the contraction limit $R\to \infty$ we have $r \to \tilde{r}/R$ where
$\tilde{r} =  \sqrt{x_2^2 + x_3^2}$ is the radial variable in the Euclidean plane.
Taking into account the limit
\begin{gather*}
\varepsilon (R) \to \tilde{\varepsilon} = \sqrt{1- \frac{\omega^2 L^2}{E^2}},
\qquad
R^2   p(R) \to \tilde{p} \equiv \frac{L^2}{E},
\end{gather*}
we get from (\ref{ELL-101}) that the equation of trajectories transforms
into the oscillator one
\begin{gather*}
\frac{x_2^2}{\tilde{B}^2} +  \frac{x_3^2}{\tilde{A}^2}  = 1,
\qquad
{\tilde{B}^2} = \frac{\tilde{p}}{1+ \tilde{\varepsilon}},
\qquad
{\tilde{A}^2} = \frac{\tilde{p}}{1 - \tilde{\varepsilon}}
\end{gather*}
The next interesting case is when $B^2 < 1 < A^2$. This conic~(\ref{ELL-001}) is neither the ellipse nor the hyperbola.
Following the paper~\cite{RANADA1} we will call this conic as the {\it ultraellipse}.  Only one semi\-axis~$b$ belongs
to the hyperbolic plane
and the next one formally is not on the real distance.  It is possible to introduce a new ``semiaxis'' $\tilde{a}$
(situated on the complex plane on the line $\tilde{a} = a+i\pi/2$) which related with the quantity~$A$ by $A^2 = \coth \tilde{a}$.
Thus, instead of~(\ref{ELL-003}) we have the conic
\begin{gather}
\label{ELL-103}
\frac{1}{\tanh^2r} =  \frac{\cos^2\varphi}{\tanh^2 b} +   {\tanh^2 \tilde{a}}{\sin^2\varphi}.
\end{gather}
There is a joint point of two conics (\ref{ELL-003}) and (\ref{ELL-103}), namely  $A^2 = 1$
($a \to \infty$ or $\tilde{a} \to \infty$). In this case the conic is given by
\begin{gather*}
\frac{1}{\tanh^2r} =  \frac{\cos^2\varphi}{\tanh^2 b} +   {\sin^2\varphi}.
\end{gather*}
This conic is an {\it equidistant curve} with equidistance~$b$ from the axis~$z_2$~\cite{RANADA1}.

Let us now consider all the possible trajectories of motion depending on the energy and angular momentum~$L^2$.

\begin{figure}[t]
\centering
\includegraphics[width=7cm,height=7cm]{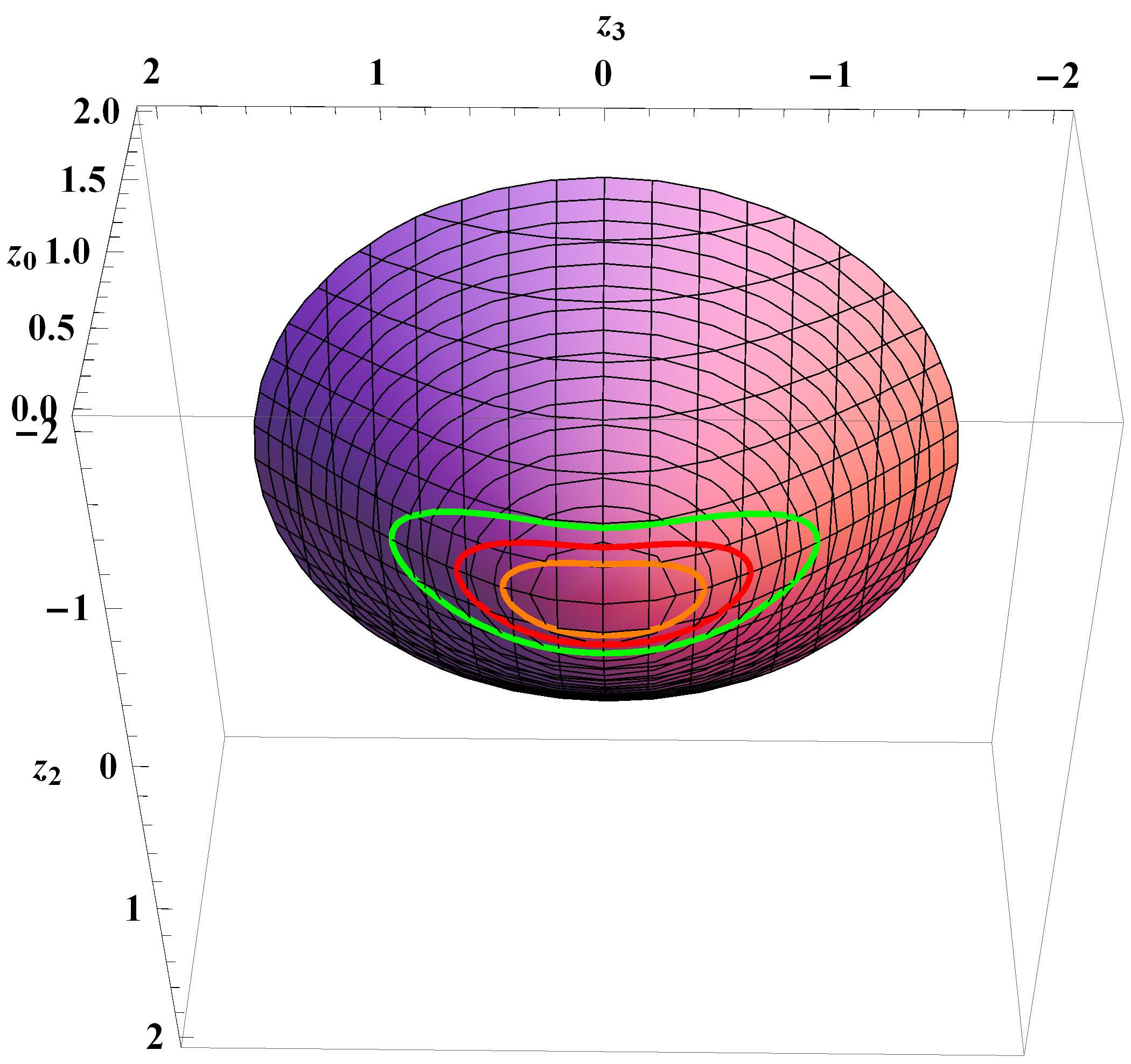}\quad
\includegraphics[width=7cm,height=7cm]{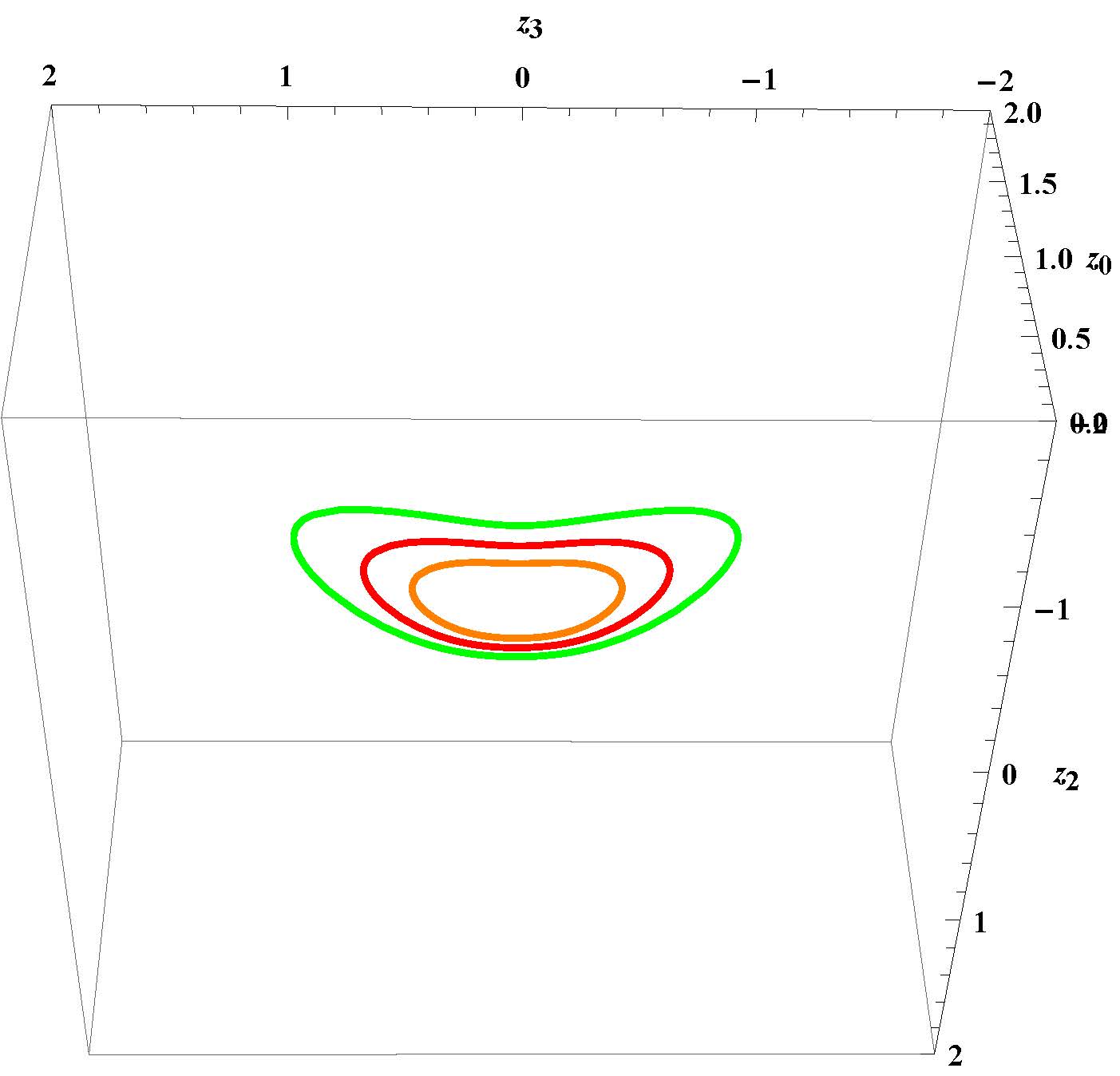}
\caption{The f\/igure shows the elliptic trajectories lying on the upper sheet of the two-sheeted hyperboloid $z_0^2-z_2^2-z_3^2 = R^2$,
$z_0 > R$ for the value $\varepsilon=0.3$ and $p=0.3, 0.4, 0.5$.}\label{Fig4}
\end{figure}

{\bf A.} First we consider the case when $U_{\text{ef\/f}}(r_0) < E < \omega^2R^2/2$ and $0 < L^2 < \omega^2R^4$.
It is clear that
\begin{gather*}
B^2 \leq A^2 =  \left\{\left(\frac{ER^2}{L^2}+\frac12\right) -
\sqrt{\left(\frac{ER^2}{L^2}+\frac12\right)^2-\frac{\omega^2 R^4}{L^2}}\right\}^{-1} < 1
\end{gather*}
and the oscillator orbits are described by the equation (\ref{ELL-003}).  Denote the minimum $b = r_{\min}$,
$(\varphi = 0)$ and maximum $a = r_{\max}$, $(\varphi = \pi/2)$ points on the orbit as a distance
from the center of  f\/ield.  From~(\ref{ELL-002}) and~(\ref{ELL-003}) we have
\begin{gather*}
\tanh^2r_{\min} =  \frac{p}{1 +  \varepsilon (R)},
\qquad
\tanh^2r_{\max} =  \frac{p}{1 -  \varepsilon (R)},
\end{gather*}
and correspondingly
\begin{gather*}
r_{\min}  =  \coth^{-1} \left\{\sqrt{\left(\frac{ER^2}{L^2}+\frac12\right)
+\sqrt{\left(\frac{ER^2}{L^2}+\frac12\right)^2-\frac{\omega^2 R^4}{L^2}}}\right\},
\\
r_{\max}  =  \coth^{-1} \left\{\sqrt{\left(\frac{ER^2}{L^2}+\frac12\right)
- \sqrt{\left(\frac{ER^2}{L^2}+\frac12\right)^2-\frac{\omega^2 R^4}{L^2}}}\right\}.
\end{gather*}
Thus we f\/ind that the trajectories of motion are {\it ellipses} lying symmetrically
to the point $z_0 = R$, $z_1 = z_2 = z_3 = 0$ on the upper sheet of the two-sheeted hyperboloid
(see Fig.~\ref{Fig4}).

\begin{figure}[t]
\centering
\includegraphics[width=7cm,height=7cm]{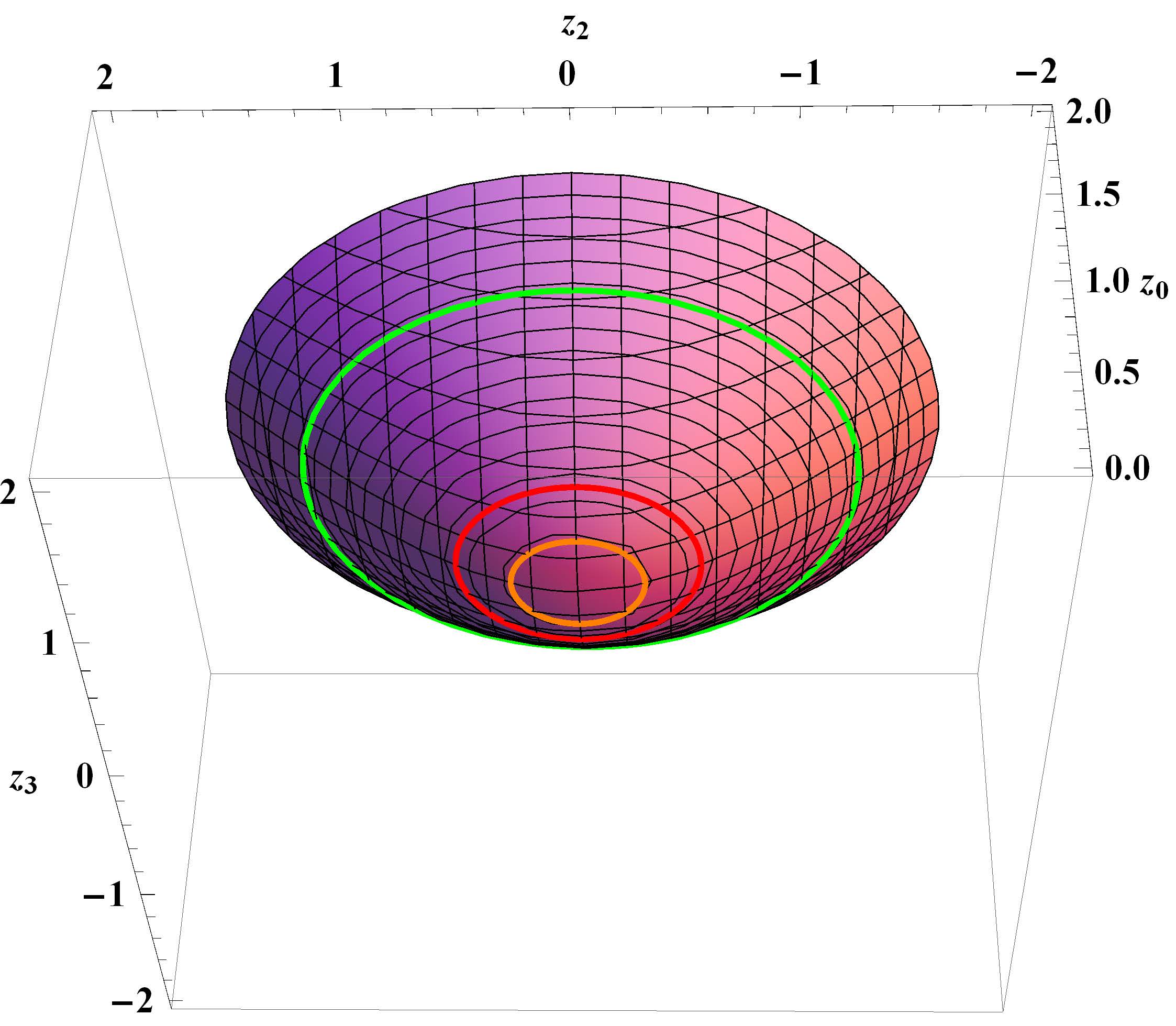}\quad
\includegraphics[width=7cm,height=7cm]{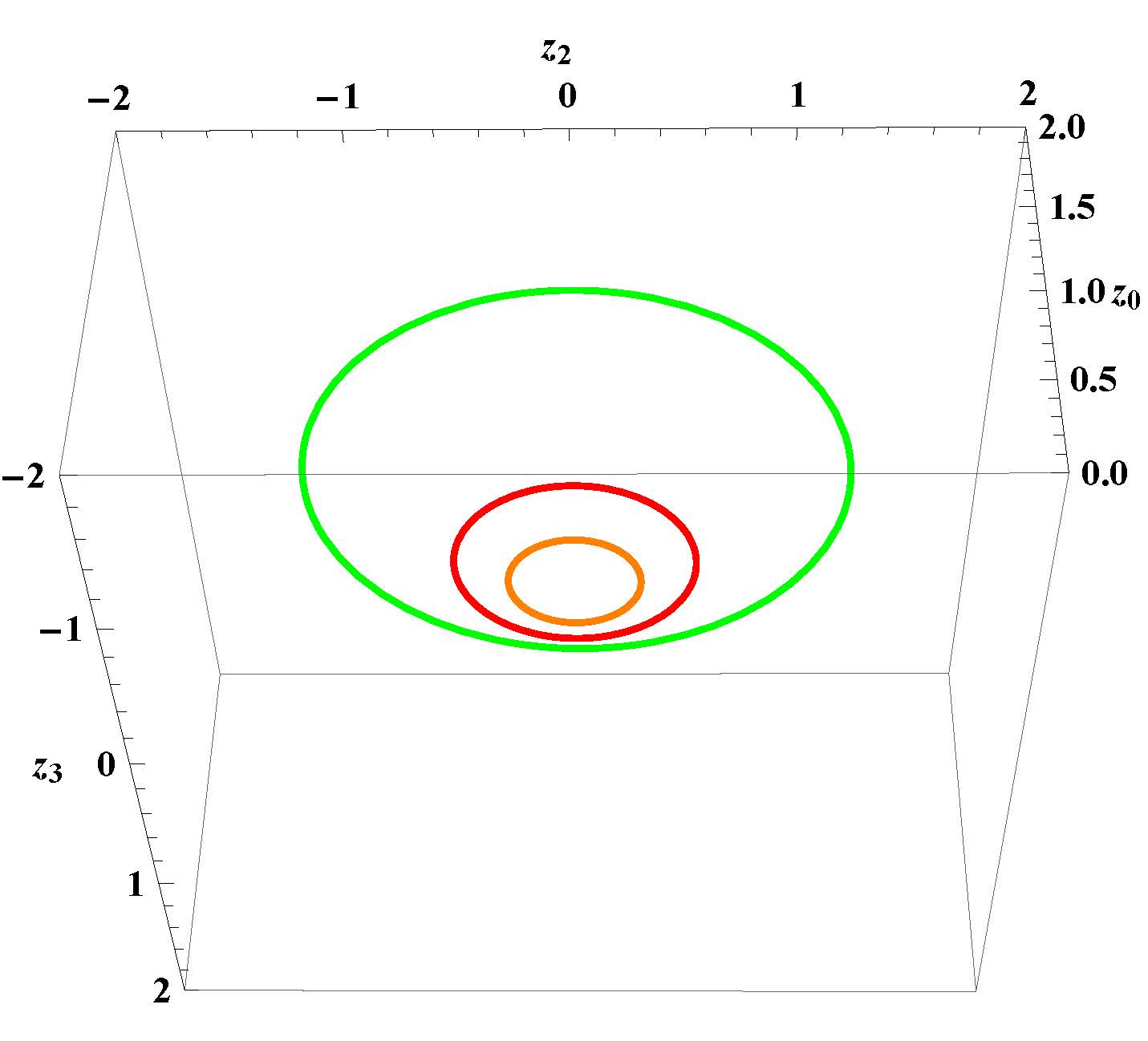}
\caption{The cyclic orbits: $\varepsilon= 0$ and $p=0.2, 0.5, 0.8$.}\label{Fig5}
\end{figure}

{\bf B.} In case of minimum energy $E = E_{\min} = U_{\text{ef\/f}}(r_0)$ we have from (\ref{H-J30}) that $\varepsilon = 0$ and
$p = \omega R^2/ \sqrt{L^2}$ and consequently $\tanh^2 r = B^2 = A^2 = \omega R^2/ \sqrt{L^2}$. Thus the orbits are circles with the radius
given by the formula (\ref{H-J10}) (see Fig.~\ref{Fig5}).

\begin{figure}[t]
\centering
\includegraphics[width=7cm,height=7cm]{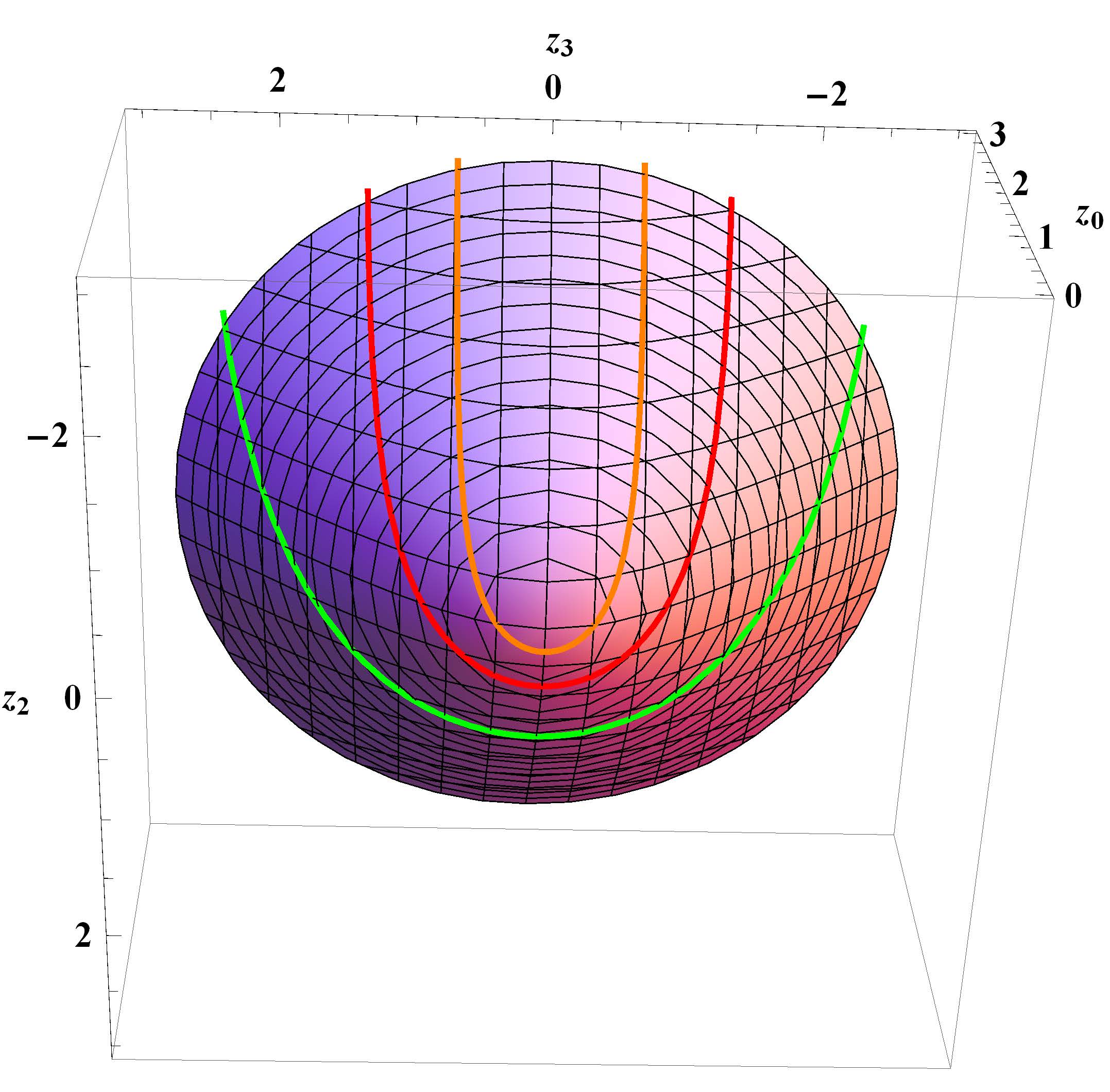}\quad
\includegraphics[width=7cm,height=7cm]{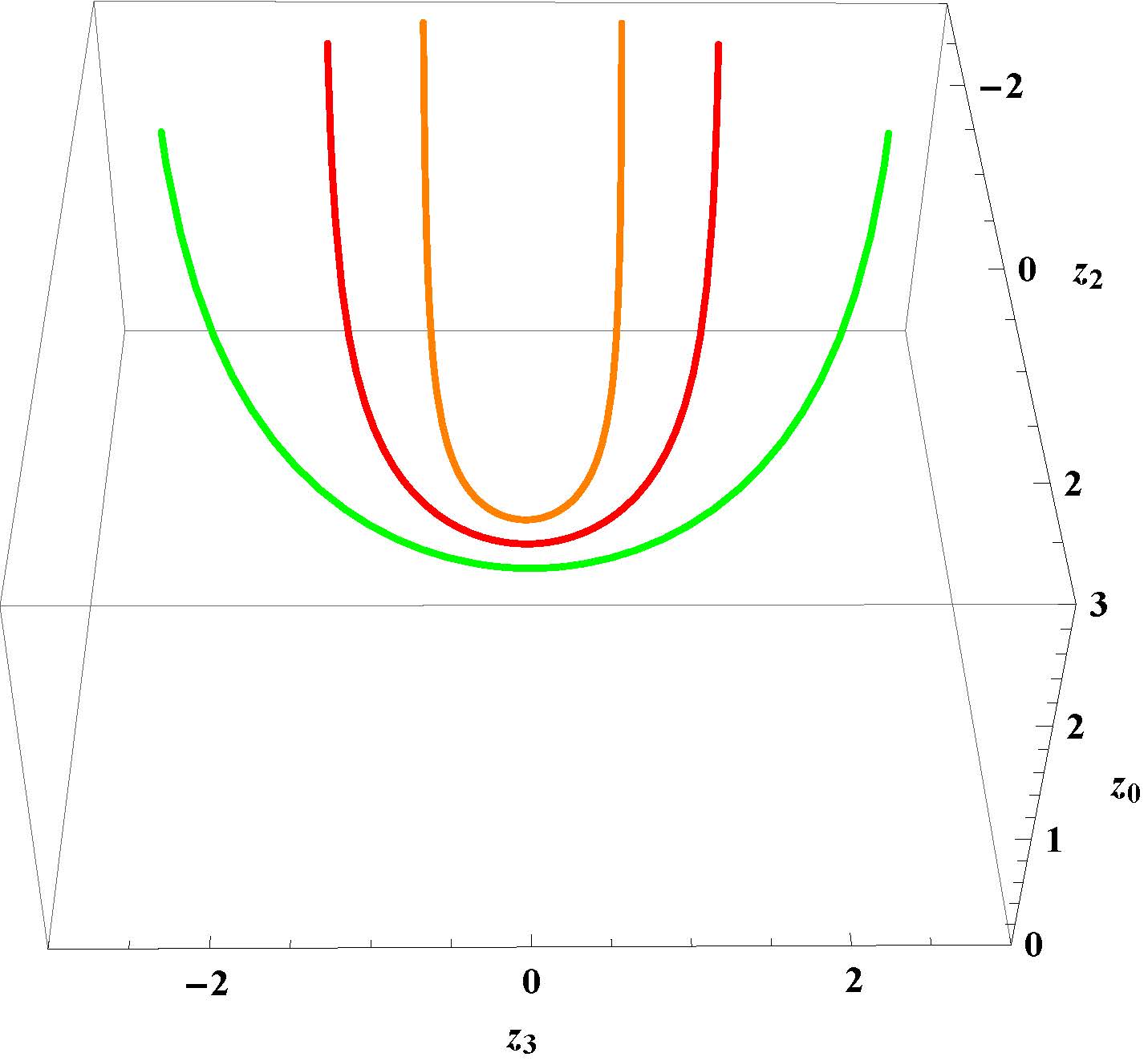}
\caption{The f\/igure shows the equidistant orbits lying on the upper sheet of the two-sheeted hyperboloid $z_0^2-z_2^2-z_3^2 = R^2$,
$z_0 > R$ with the value of pairs $(p, \varepsilon) = (1/3, 2/3);$ $(2/3, 1/3); (8/9, 1/9)$.}\label{Fig6}
\end{figure}

{\bf C.} For the case of energy values $E = \omega^2R^2/2$ we get that
\begin{gather*}
p (R)  = \frac{2A}{\omega^2 R^4 + L^2},
\qquad
\varepsilon (R) =   \frac{|\omega^2 R^4 - L^2|}{\omega^2 R^4 + L^2},
\end{gather*}
therefore for $0 < L^2 < \omega^2 R^4$  we get $B^2 = L^2/\omega^2 R^4 < 1$ and $A^2 = 1$.
The conic is
\begin{gather*}
\label{ELL-204}
\frac{1}{\tanh^2r} =   \frac{\omega^2 R^4}{L^2}  \cos^2\varphi  + \sin^2\varphi,
\end{gather*}
which represents the equidistant curves (see Fig.~\ref{Fig6}).
The minimal distance $r_{\min}$ from the center is given by the formula
\begin{gather*}
r_{\min} = \coth^{-1} \left(\frac{\omega R^2}{\sqrt{L^2}}\right).
\end{gather*}
Let $L^2 = \omega^2 R^4$. Then $B^2 = A^2 = 1$ and the conic is a ``largest'' circle with
radius $r = \infty$.  For the case $L^2 > \omega^2 R^4$ we obtain that $B^2 = 1$,
$A^2 = L^2/\omega^2 R^4 >1$. Then from the formula~(\ref{ELL-001}) it follows that $\tanh r >1$
and no any oscillator orbits exist.

\begin{figure}[t]
\centering
\includegraphics[width=7cm,height=7cm]{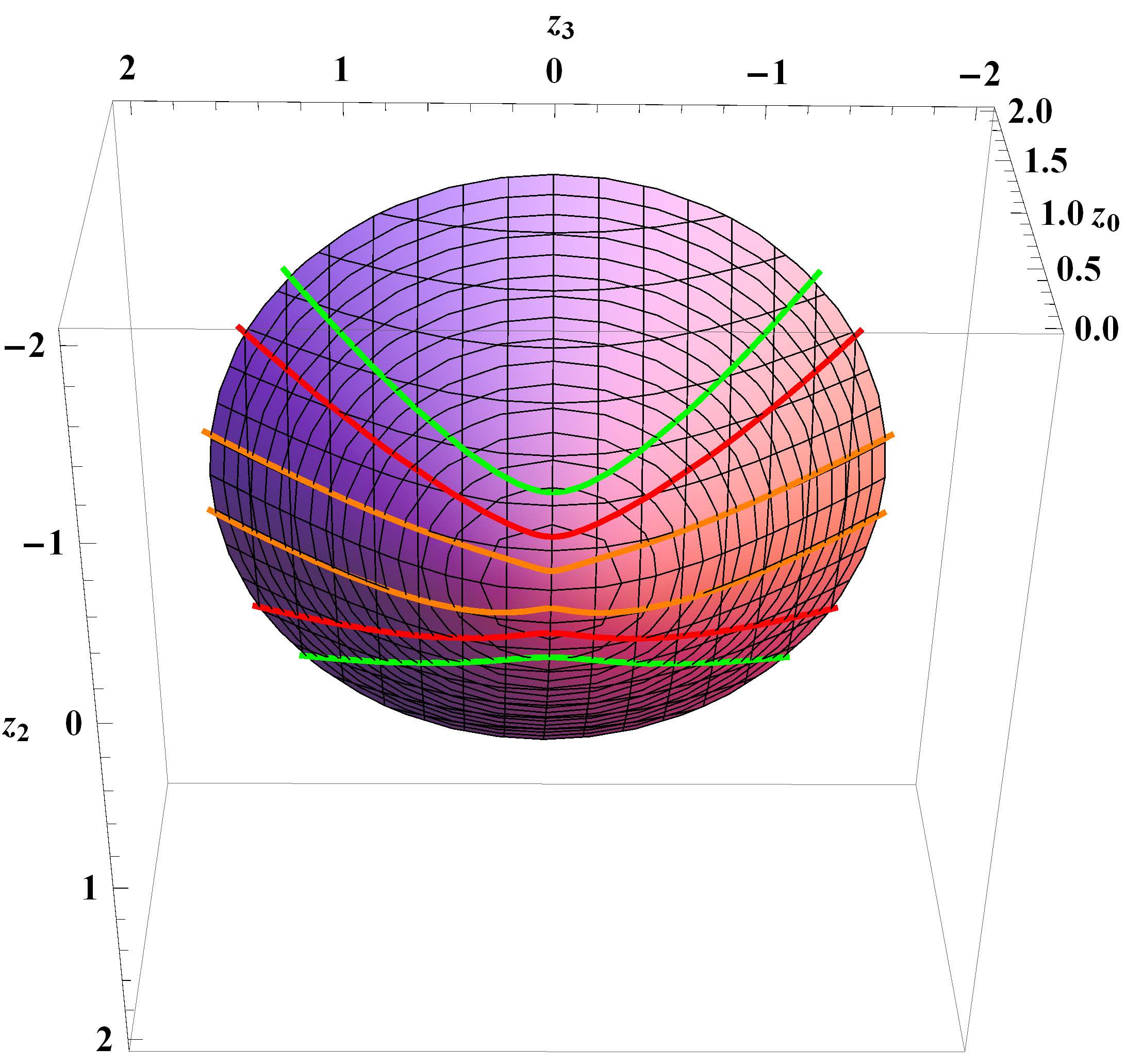}\quad
\includegraphics[width=7cm,height=7cm]{Fig7a}
\caption{The f\/igure shows the ultraellipses lying on the upper sheet of the two-sheeted hyperboloid $z_0^2-z_2^2-z_3^2 = R^2$,
$z_0>R$ for the value $\epsilon=0.8$ and $p=0.2, 0.5, 0.8$.}\label{Fig7}
\end{figure}

{\bf D.}
For the energy $E > \omega^2R^2/2$ it is easy to see that for any positive $L^2 > 0$
\begin{gather*}
A^2 = \left\{\left(\frac{ER^2}{L^2}+\frac12\right)- \sqrt{\left(\frac{ER^2}{L^2}+\frac12\right)^2 -
\frac{\omega^2 R^4}{L^2}}\right\}^{-1} > 1,
\qquad
B^2 < 1.
\end{gather*}
The motion of a particle is determined by the equation (\ref{ELL-103}) where $\tanh^2 \tilde{a} = 1/A^2$.
The trajectories are ultraellipses and describe the motion of a particle from the minimum point~$r_{\min}$:
\begin{gather*}
r_{\min}  =  \coth^{-1} \left\{\sqrt{\left(\frac{ER^2}{L^2}+\frac12\right)+
\sqrt{\left(\frac{ER^2}{L^2}+\frac12\right)^2 - \frac{\omega^2 R^4}{L^2}}}\right\},
\end{gather*}
to inf\/inity (see Fig.~\ref{Fig7}). On the other hand side $B^2 \cdot A^2 = \omega^2 R^4/ L^2$,
so that for $L^2 < \omega^2 R^4$ we get $1/A^2 < B^2 < 1$, whereas for $L^2 > \omega^2 R^4$ follows that
$B^2 < 1/A^2 < 1$ and the value of $L^2=\omega^2 R^4$ or $B^2 = 1/A^2$ separates two set of ultraellipses.

Let us also note that the in contraction limit $R\to \infty$ these orbits corresponds to the
Euclidean oscillator orbits with the large values of energy (the straight line $x_2^2 = {\tilde B}^2$).

\section[The trajectories for $L^2\leq 0$]{The trajectories for $\boldsymbol{L^2\leq 0}$}

To simplify further formulas we set f\/irst $p_\varphi = 0$. Then, from equation~(\ref{H-J0-23}) it follows that
the motion occurs at a constant value of the azimutal angle $\varphi = \varphi_0$ that is limited by the condition
$z_3/z_2 = \tan\varphi_0$. To further simplify it is enough to choose $\varphi_0 = 0$ or $\varphi_0 = \pi$.
Thus we get that trajectory of the motion lies on the one-sheeted hyperboloid
$z_0^2+z_1^2-z_2^2=R^2$.
The formula~(\ref{H-J16-0}) gives us the equation of the trajectory in the region $z_0 > R$:
\begin{gather}
\label{TRAJECTORY-01}
\coth^2 r = \left(\frac{1}{2} - \frac{E R^2}{|L^2|}\right) +
\sqrt{\left(\frac{1}{2}  - \frac{E R^2}{|L^2|}\right)^2 + \frac{\omega^2R^4}{|L^2|}} \cosh\big(2 \tau + 4\sqrt{|L^2|}\beta\big).
\end{gather}
Performing  the further transformation  $r \to i\chi$ and $\tau \to \mu - i\pi/2$ in formula (\ref{TRAJECTORY-01}), we
obtain the equation of the trajectory in the region $0 < z_0 < R$:
\begin{gather}
\label{TRAJECTORY-COM.01}
\cot^2\chi = - \left(\frac{1}{2} - \frac{E R^2}{|L^2|}\right) +
\sqrt{\left(\frac{1}{2}  - \frac{E R^2}{|L^2|}\right)^2 + \frac{\omega^2R^4}{|L^2|}} \cosh\big(2 \mu + 4\sqrt{|L^2|}\beta\big).
\end{gather}
In the formula of trajectory~(\ref{TRAJECTORY-01}) we must distinguish two cases, namely for the value of energy $E < \omega^2 R^2/2$
and $E \geq \omega^2 R^2/2$.

In the f\/irst case $E < \omega^2 R^2/2 $ from equation~(\ref{TRAJECTORY-01}) it follows that for any value
of the variable $\tau \in (-\infty, \infty)$ we have that $\coth r > 1$. Therefore, the trajectory of the motion extends from the point
$r=0$ at the $\tau \to - \infty$ $(z_0 = R$, $z_1 <0$, $z_2 >0)$ to its maximum
\begin{gather*}
r_{\max}  = \coth^{-1}\sqrt{\left(\frac{1}{2} - \frac{E R^2}{|L^2|}\right) +
\sqrt{\left(\frac{1}{2} - \frac{E R^2}{|L^2|}\right)^2 + \frac{\omega^2R^4}{|L^2|}}},
\end{gather*}
at the point $\tau = - 2\sqrt{|L^2|}\beta$ and then goes back to the point $r=0$ when $\tau \to \infty$ $(z_0 = R$, $z_1 > 0$, $z_2 >0)$.
Further on, the particle penetrates through the point $z_0 = R$ from the region $z_0 > R$ to the region $0 < z_0 < R$,
which, as it follows from the equation~(\ref{TRAJECTORY-COM.01}), corresponds to the value of angles $\mu \to \infty$ and
$\chi \to  0$, $(z_0 < R$, $z_1 > 0$, $z_2 >0)$. Further trajectory extends to the maximal value~$\chi_{\max}$:
\begin{gather*}
\chi_{\max}  = \cot^{-1}\sqrt{- \left(\frac{1}{2} - \frac{E R^2}{|L^2|}\right) +
\sqrt{\left(\frac{1}{2} - \frac{E R^2}{|L^2|}\right)^2 + \frac{\omega^2R^4}{|L^2|}}} \leq \frac{\pi}{2},
\end{gather*}
at the point $\mu = - 2\sqrt{|L^2|}\beta$, and then continue to $\mu \to - \infty$, $\chi \to 0$ $(z_0 < R$, $z_1 > 0$, $z_2 < 0)$.
After, the particle again passes the point $z_0 = R$ and penetrates to the region $z_0 \geq R$.
Further using similar reasoning it can be shown that the trajectories in case of $E < \omega^2 R^2/2$,
are a closed curve lying on the one-sheeted hyperboloid $z_0^2+z_1^2-z_2^2=R^2$, $z_0 > 0$, so the motions are bounded
and periodic. The same situation takes place for the case of $z_0 <0$.

\begin{figure}[t]
\centering
\includegraphics[width=7cm,height=7cm]{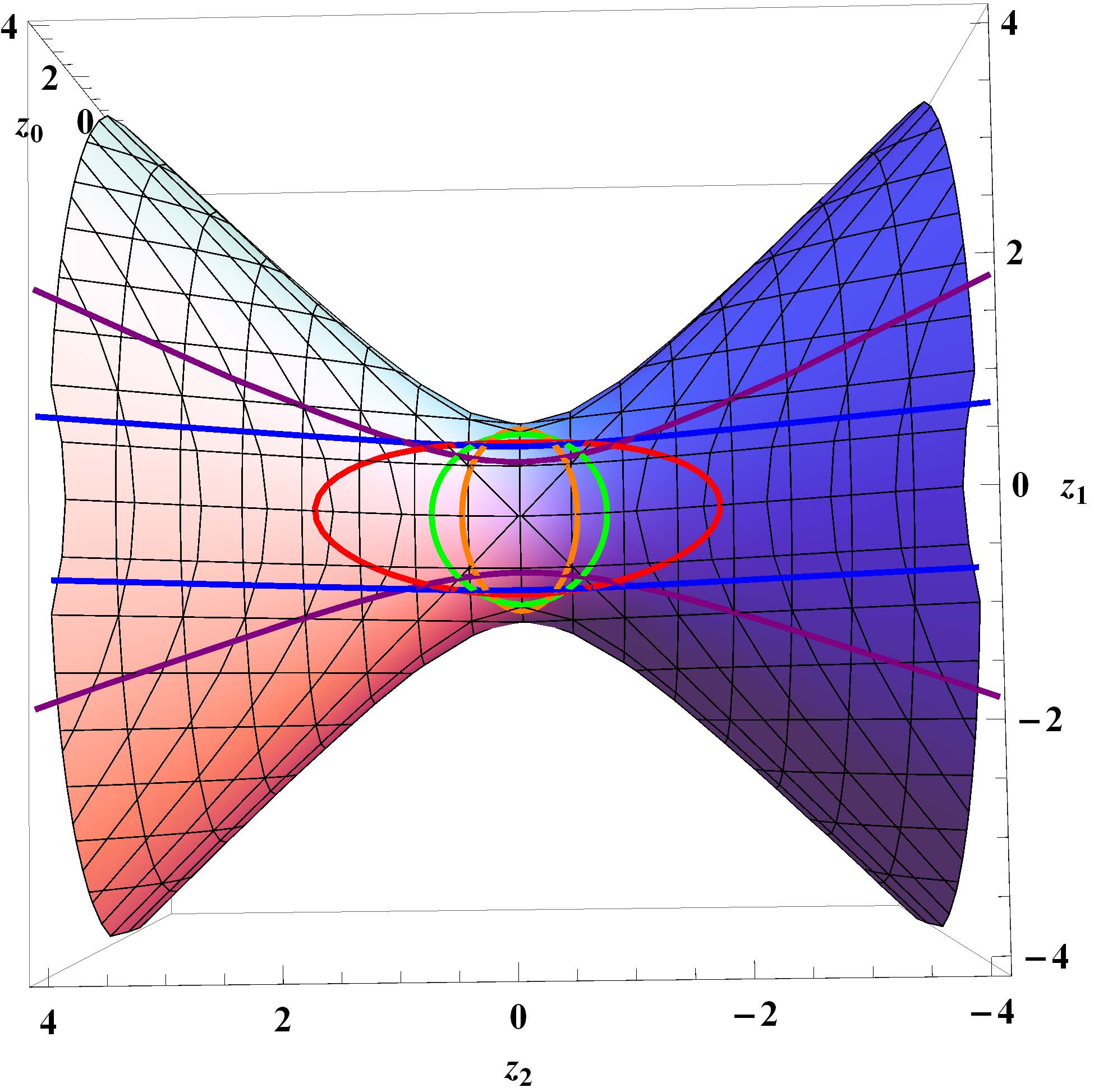}\quad
\includegraphics[width=7cm,height=7cm]{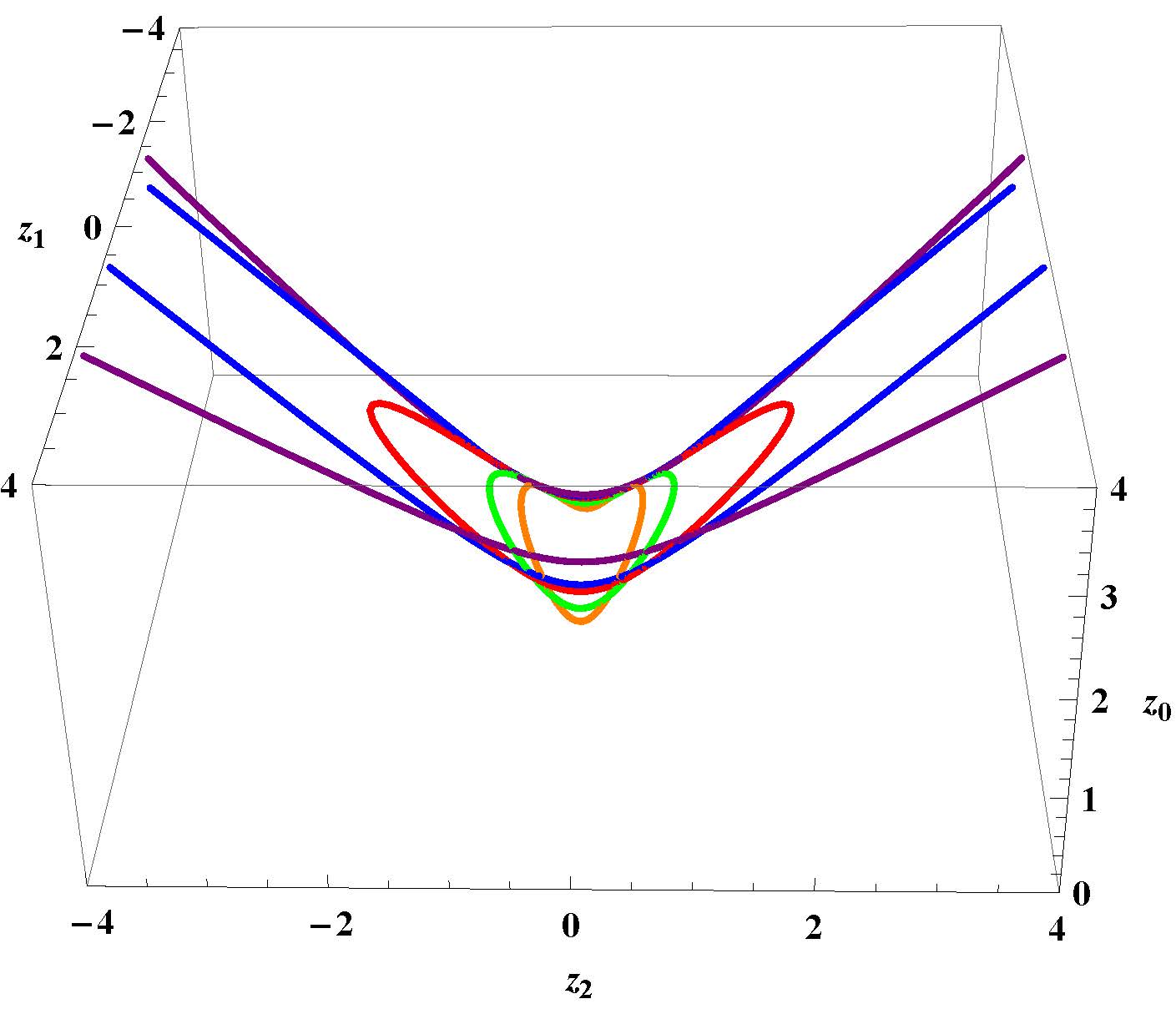}
\caption{The trajectories of motion in the case of $|L^2| = 1$; $E = -3/2, -1/2, 1/4, 1/2, 3/2$; $\omega = R = 1$.}\label{Fig8}
\end{figure}

In the case of $E \geq \omega^2 R^2/2$ it is easy to see that the inequality
\begin{gather*}
\sqrt{\left(\frac{1}{2}  - \frac{E R^2}{|A|}\right)^2 + \frac{\omega^2R^4}{|L^2|}} \leq \frac{1}{2} + \frac{E R^2}{|L^2|}
\end{gather*}
is valid. Thus the trajectory of the motion, depending on the sign of variable $\tau$ is split into two paths.
One of the paths begins from the large~$r$ at the minimal point
\begin{gather*}
\tau_{\min}  = - 2\sqrt{|L^2|}\beta + \frac{1}{2} \cosh^{-1} \frac{\left(\frac{1}{2} + \frac{E R^2}{|L^2|}\right)}
{\sqrt{\left(\frac{1}{2}  - \frac{E R^2}{|L^2|}\right)^2 + \frac{\omega^2R^4}{|L^2|}}}.
\end{gather*}
and continues to the point $r=0$ at $\tau \to \infty$ $(z_0 = R$, $z_2 > 0)$. Then the trajectory passing the part of
$0 < z_0 < R$ goes back from $(z_0 = R$, $z_2 < 0)$ at the point $r=0$, $\tau \sim \infty$ to
$r \in \infty$ at~$\tau_{\min}$. The second path is symmetric with respect to axis $z_1$.
Thus the trajectories of motion in the case of $E \geq \omega^2 R^2/2$ are not bounded.
Some examples of trajectories for the f\/ixed negati\-ve~$L^2$ and various values of energy~$E$,
are presented on the Fig.~\ref{Fig8}.

\begin{figure}[t]
\centering
\includegraphics[width=7cm,height=7cm]{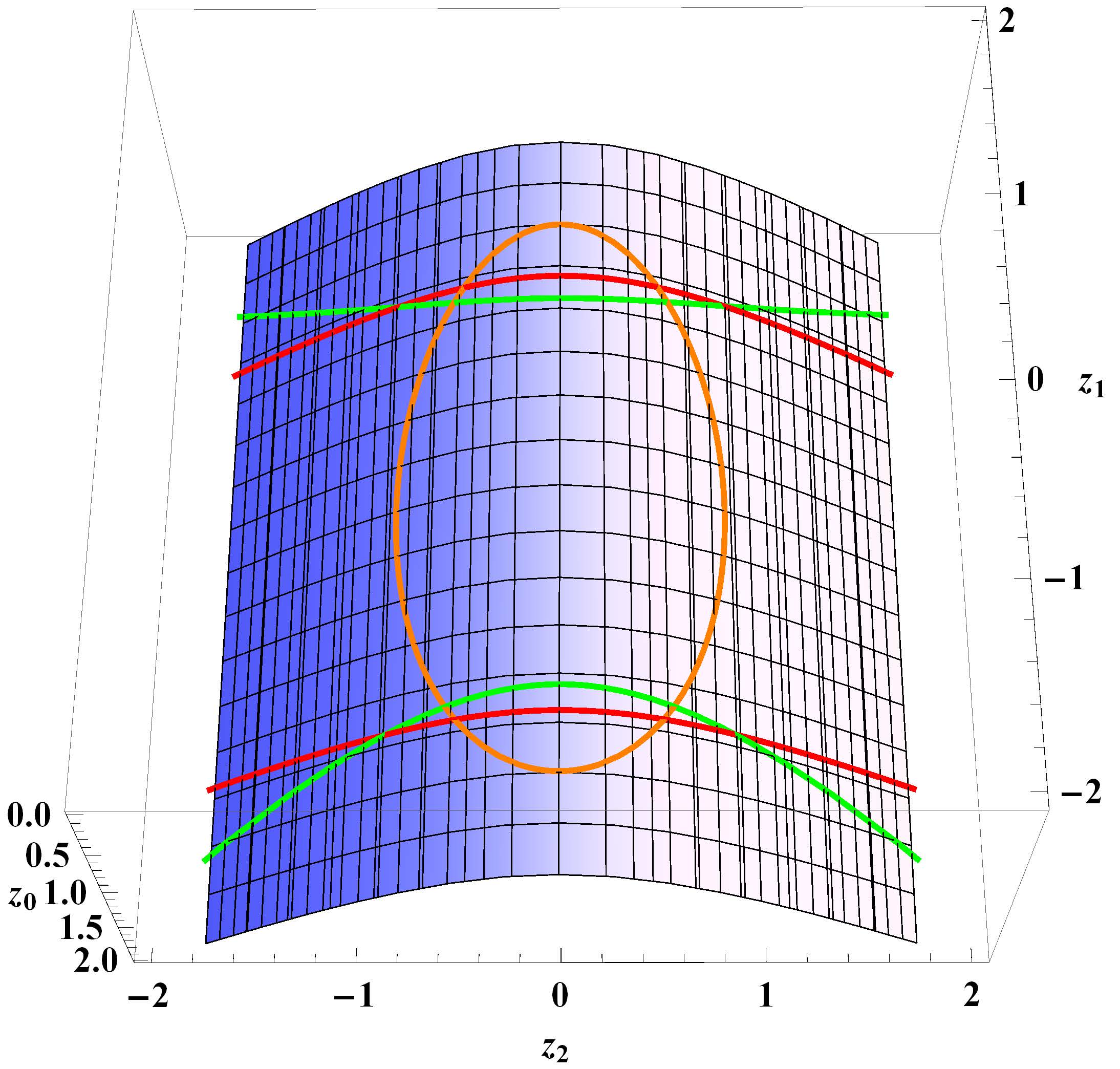}\quad
\includegraphics[width=7cm,height=7cm]{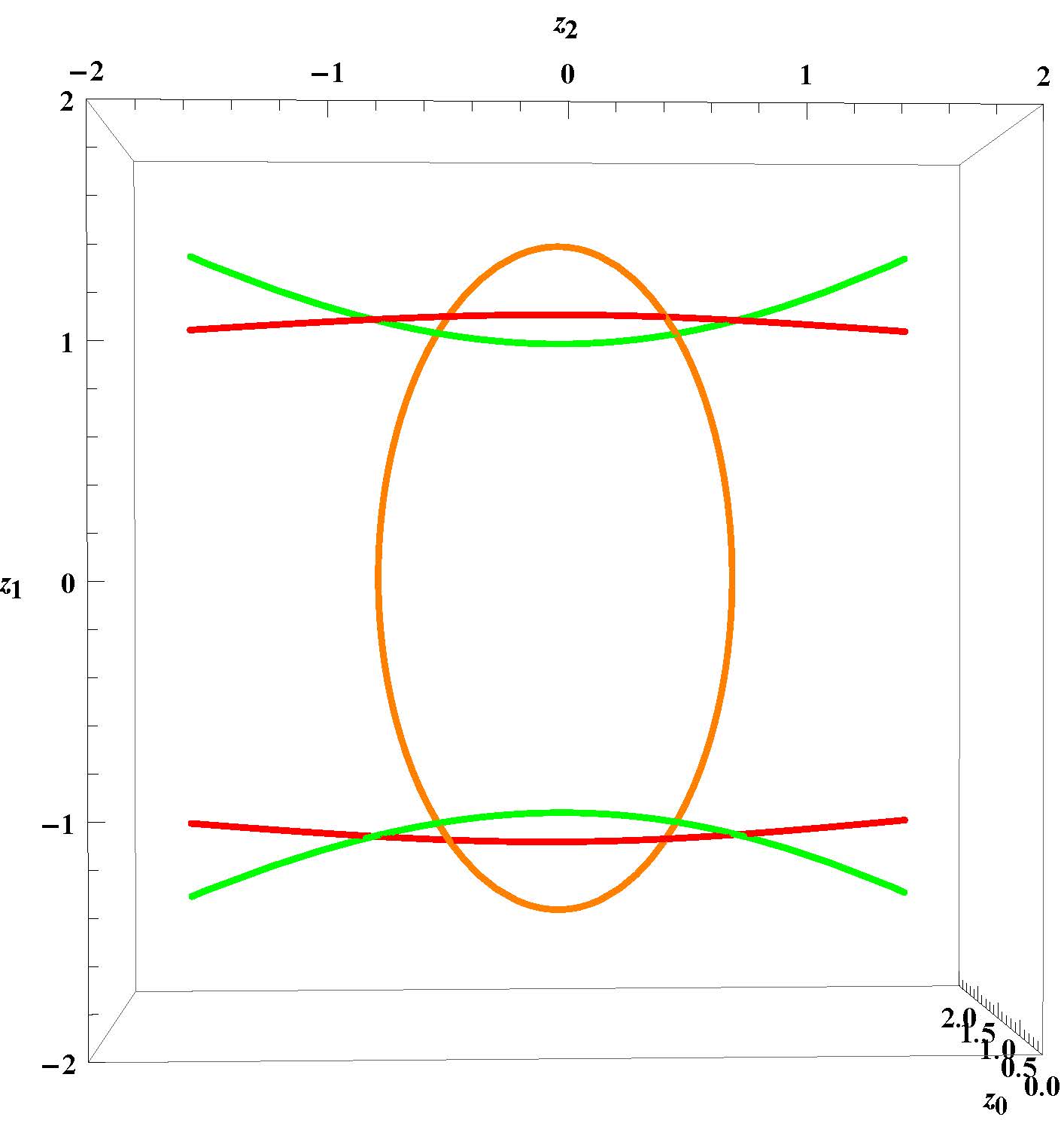}
\caption{The bounded and inf\/inite trajectories of the motion for $L^2=0$ lying on the hyperbolic cylinder
$z_0^2 - z_2^2 = R^2$, $z_1^2 = z_3^2$, and $z_0 \geq R$.
The f\/igure shows the cases $E=0.2,  0.5,  0.8$; $\omega = R = p_{\varphi} = 1$.}\label{Fig9}
\end{figure}

In the case $L^2=0$ it is easy to get from (\ref{H-J23-A0})
\begin{gather*}
\label{H-J28-3}
\coth^2 r =  \frac{\omega^2R^2}{2 E} + R \sqrt{E} \left(2\beta - \tan \varphi/p_\varphi \right)^2,
\end{gather*}
with $\varphi_0 = 0$.
In the case of $E < \omega^2R^2/2$ the bounded motion takes place $r_{\min} = 0$ ($\varphi = \pi/2$) and
$r_{\max} = \coth^{-1}\sqrt{\frac{\omega^2R^2}{2 E}}$ ($\varphi = \arctan{2\beta p_\varphi}$),
whereas for the $E \geq \omega^2R^2/2$ the orbits are inf\/inite: $r \in [0, \infty)$. The trajectories of
the motion can be presented on the hyperbolic cylinder $z_0^2 - z_2^2 = R^2$, $z_1^2 = z_3^2$, $z_0 \geq R$ (see Fig.~\ref{Fig9}).

\section{Conclusion}

We have shown that the notion of harmonic oscillator problem can be extended not only to the sphere and two-sheeted hyperboloid
but also to the hyperbolic space~$H_2^2$. It was proved that the harmonic oscillator problem on $H_2^2$ is exactly solvable and
also belongs  to the class of superintegrable systems. We have constructed the dynamical algebra of symmetry for this system,
which is nonlinear and quadratic (so-called Higgs algebra). We completely solved the Hamilton--Jacobi equation for harmonic oscillator
problem in the geodesic pseudo-spherical systems of coordinates.
It was shown that for positive value of  the Lorentzian momentum $L^2 > 0$ all trajectories of motion lie on the upper (or lower)
sheets of two dimensional two-sheeted hyperboloid $z_0^2-z_2^2-z_3^2=R^2$. These trajectories are always conics centered in the origin
of potential $r=0$. For the special values of energy $E_{\min} < E < \omega^2 R^2/2$ and momentum $L^2 < \omega^2 R^4$ all the orbits
are ellipses (or circles for $E=E_{\min})$.  In case when $E > \omega^2 R^2/2$ independently of the value of $L^2$, the oscillator
orbits are ultraellipses or equidistant curves for $E = \omega^2 R^2/2$. We have seen that in case of negative values of Loreinzian
momentum $L^2 \leq 0$ the oscillator orbits lie on the one-sheeted hyperboloid $z_0^2+z_1^2-z_2^2=R^2$ and are bounded and periodic
for $E < \omega^2 R^2/2 $ and inf\/inite for $E \geq \omega^2 R^2/2$. The similar situation is valid for $L^2=0$, but in this case the
orbits lie on the hyperbolic cylinder $z_0^2 - z_2^2 = R^2$, $z_1^2 = z_3^2$.

Let us make short comments concerning the connection of the classical and quantum case.
The quantum-mechanical counterpart of the angular momentum operator~(\ref{classical1})
comes through the replacement $p_\mu \to - i \partial/\partial z_\mu$ and is given by
\begin{gather*}
\hat{L}_1= - i(z_2\partial_3-z_3\partial_2),
\qquad
\hat{L}_2 = -i(z_1\partial_3+z_3\partial_1),
\qquad
\hat{L}_3 = i (z_1\partial_2+z_2\partial_1).
\end{gather*}
Then in the pseudo-spherical coordinates (\ref{coor:1}) the operator $\hat{L}^2$ takes the form
\begin{gather*}
\label{ANGULAR-QUANTUM2}
\hat{L}^2 = \hat{L}_1^2 - \hat{L}_2^2 - \hat{L}_3^2 = \left(\frac{1}{\cosh\tau}\frac{\partial}{\partial \tau}
\cosh\tau\frac{\partial}{\partial \tau}-\frac{1}{\cosh^2\tau} \frac{\partial^2}{\partial\varphi^2}\right),
\end{gather*}
and coincide with the Casimir operator of ${\rm SO}(2,1)$ group. Thus the Schr\"odinger equation for the harmonic oscillator
potential can be written as
\begin{gather}
\label{ANGULAR-QUANTUM3}
\frac{1}{\sinh^2r}\frac{\partial}{\partial r}
\sinh^2r\frac{\partial\Psi}{\partial r} + \left[2 R^2 E  - \frac{\hat{L}^2}{\sinh^2r}
-  \omega^2 R^4 \tanh^2r \right] \Psi = 0,
\end{gather}
and solved by separation of variables via the ansatz $\Psi(r, \tau, \varphi) = {\cal R}(r) {\cal Y}(\tau, \varphi)$.
The pseudo-spherical function ${\cal Y}$ is a eigenfunction of operator ${\hat{L}^2}{\cal Y} = \ell(\ell+1){\cal Y}$ which describes
the quantum geodesic motion on the two-dimensional one-sheeted hyperboloid. The spectrum of $\ell$ can take as well as the real values:
$\ell = 0,1,\dots$ (discrete series of representation of ${\rm SO}(2,1)$ group) and complex value $\ell = -1/2 + i \rho$, $\rho>0$ (continuous
principal series). In the f\/irst case the eigenvalue of ${\hat{L}^2}$ operator is positive and in the second one negative.
The exact solution of the Schr\"odinger equation~(\ref{ANGULAR-QUANTUM3}) for the positive eigenvalues of operator $\hat{L}^2$ has been
constructed in the previous paper~\cite{PETPOG3}. It was shown that as in the case of two-sheeted hyperboloid, the energy spectrum contains
the scattering states and a f\/inite number of  degenerate bound states. This fact coincides  with the existence of closed and inf\/inite orbits
for positive~$L^2$ in classical case. We have not considered in the article~\cite{PETPOG3} the quantum motion in the case of negative eigenvalue of~$\hat{L}^2$ because of the strong singularity at the center of harmonic oscillator potential, although it is clear that the system has a
discrete spectrum. This work is in progress.

Finally, we wish to emphasize that the Kepler--Coulomb and harmonic oscillator potentials are the ``building block'' upon which most of
superintegrable potentials can be constructed. Thus the investigation of these systems is important for the further study and understanding
of more complicated superintegrable systems in the hyperbolic space~$H_2^2$.

\appendix

\section{Symmetry algebra}\label{appendixA}

The nonvanishing Poisson brackets between the components of Demkov--Fradkin tensor ${\cal D}_{ij}$
and~${\cal L}_i$:
\begin{alignat*}{4}
& \{{\cal D}_{12}, {\cal L}_1\} = -{\cal D}_{13}, \qquad         &&\{{\cal D}_{12}, {\cal L}_2\}  = - {\cal D}_{23}, \qquad &&
 \{{\cal D}_{12}, {\cal L}_3\} =-{\cal D}_{11}-D_{22}, &
\\
& \{{\cal D}_{13}, {\cal L}_1\} =  {\cal D}_{12}, \qquad && \{{\cal D}_{13}, {\cal L}_2\} = - {\cal D}_{11} - {\cal D}_{33}, \qquad &&
 \{{\cal D}_{13}, {\cal L}_3\}] = -{\cal D}_{23}, &
\\
& \{{\cal D}_{23}, {\cal L}_1\} ={\cal D}_{22}-{\cal D}_{33}, \qquad & &\{{\cal D}_{23}, {\cal L}_2\} = - {\cal D}_{12}, \qquad &&
 \{{\cal D}_{23}, {\cal L}_3\} = -{\cal D}_{13}, &
\\
& \{{\cal D}_{11}, {\cal L}_2\} = -2 {\cal D}_{13}, \qquad &&\{{\cal D}_{11}, {\cal L}_3\} = -2{\cal D}_{12}, \qquad &&
 \{{\cal D}_{22}, {\cal L}_1\} = -2{\cal D}_{23}, &
\\
& \{{\cal D}_{22}, {\cal L}_3] = -2{\cal D}_{12}, \qquad & &\{{\cal D}_{33}, {\cal L}_1] =2{\cal D}_{23}, \qquad &&
 \{{\cal D}_{33}, {\cal L}_2\} =- 2{\cal D}_{13},&
\end{alignat*}
The same between ${\cal D}_{ik}$:
\begin{gather*}
\{{\cal D}_{11}, {\cal D}_{12}\} =2\omega^2{\cal L}_3 + \frac{2}{R^2} {\cal L}_3 {\cal D}_{11},
\qquad   \{{\cal D}_{11}, {\cal D}_{13}\} =2\omega^2{\cal L}_2+\frac{2}{R^2}{\cal L}_2 {\cal D}_{11},
\\
\{{\cal D}_{11}, {\cal D}_{23}\} =\frac{2}{R^2}({\cal L}_2 {\cal D}_{12} + {\cal L}_3 {\cal D}_{13}),
\qquad    \{{\cal D}_{11}, {\cal D}_{22}\} =\frac{4}{R^2} {\cal L}_3 {\cal D}_{12},
\\
\{{\cal D}_{22}, {\cal D}_{12}\} =2\omega^2{\cal L}_3 - \frac{2}{R^2} {\cal L}_3 {\cal D}_{22},
\qquad
 \{{\cal D}_{22}, {\cal D}_{13}\} =-\frac{2}{R^2}({\cal L}_3 {\cal D}_{23}+ {\cal L}_1 {\cal D}_{12}),
\\
\{{\cal D}_{22}, {\cal D}_{23}\} =2\omega^2{\cal L}_1 - \frac{2}{R^2} {\cal L}_1 {\cal D}_{22},
\qquad
 \{{\cal D}_{22}, {\cal D}_{33}\} = -\frac{4}{R^2} {\cal L}_1 {\cal D}_{23},
\\
\{{\cal D}_{33}, {\cal D}_{12}\} =-\frac{2}{R^2}({\cal L}_2 {\cal D}_{23} - {\cal L}_1 {\cal D}_{13}),\qquad
 \{{\cal D}_{33}, {\cal D}_{13}\} =2\omega^2{\cal L}_2 - \frac{2}{R^2} {\cal L}_2 {\cal D}_{33},
\\
\{{\cal D}_{33}, {\cal D}_{23}\} =-2\omega^2{\cal L}_1 + \frac{2}{R^2} {\cal L}_1 {\cal D}_{33},\qquad
 \{{\cal D}_{33}, {\cal D}_{11}\} =-\frac{4}{R^2} {\cal L}_2 {\cal D}_{13},
\\
\{{\cal D}_{12}, {\cal D}_{13} \}
 =  - \left(2\omega^2-\frac{1}{4R^4}\right) {\cal L}_1+\frac{1}{R^2}
\left(  {\cal L}_1 {\cal D}_{11} +   {\cal L}_2 {\cal D}_{12} +   {\cal L}_3 {\cal D}_{13} \right),
\\
\{{\cal D}_{12}, {\cal D}_{23}\}
 =  \left(2\omega^2-\frac{1}{4R^4}\right)L_2+\frac{1}{R^2}
\left( {\cal L}_1 {\cal D}_{12} +   {\cal L}_2 {\cal D}_{22} -  {\cal L}_3 {\cal D}_{23}\right),
\\
\{{\cal D}_{13}, {\cal D}_{23} \}
 =  - \left(2\omega^2-\frac{1}{4R^4}\right)  {\cal L}_3+
\frac{1}{R^2} \left(-  {\cal L}_1 {\cal D}_{13} +   {\cal L}_2 {\cal D}_{23} -  {\cal L}_3 {\cal D}_{33}\right).
\end{gather*}

\subsection*{Acknowledgments}

The work of G.P.~was partially supported under the Armenian-Belarus grant Nr.~13RB-035 and Armenian national grant Nr.~13-1C288.

\pdfbookmark[1]{References}{ref}
\LastPageEnding

\end{document}